# SOFT BILLIARDS WITH CORNERS

D. TURAEV AND V. ROM-KEDAR

ABSTRACT. We develop a framework for dealing with smooth approximations to billiards with corners in the two-dimensional setting. Let a polygonal trajectory in a billiard start and end up at the same billiard's corner point. We prove that smooth Hamiltonian flows which limit to this billiard have a nearby periodic orbit if and only if the polygon angles at the corner are "acceptable". The criterion for a corner polygon to be acceptable depends on the smooth potential behavior at the corners, which is expressed in terms of a *scattering function*. We define such an asymptotic scattering function and prove the existence of it, explain how can it be calculated and predict some of its properties. In particular, we show that it is non-monotone for some potentials in some phase space regions. We prove that when the smooth system has a limiting periodic orbit it is hyperbolic provided the scattering function is not extremal there. We then prove that if the scattering function is extremal, the smooth system has elliptic periodic orbits limiting to the corner polygon, and, furthermore, that the return map near these periodic orbits is conjugate to a small perturbation of the Hénon map and therefore has elliptic islands. We find from the scaling that the island size is typically algebraic in the smoothing parameter and exponentially small in the number of reflections of the polygon orbit.

## 1. INTRODUCTION

Modelling Hamiltonians with steep potentials as singular, billiard-like systems has proved to be a useful concept in a variety of applications (cold atoms motion [9], molecular dynamics [5], [8], fundamentals of statistical physics [15],[18], semiclassical approximations of particles motion [17] and others). It is natural to ask what are the conditions under which such an approximation is justified (i.e. the limit is regular), and to develop tools for analyzing new dynamical effects which appear when the approximation fails, see [12].

The simplest setting at which these issues arise is represented by two dimensional billiard domains, i.e. when one studies the behavior of smooth two degrees of freedom Hamiltonian systems:

$$H = \frac{1}{2}(p_x^2 + p_y^2) + V(x,y;\varepsilon), \tag{1.1}$$

which limit, as $\varepsilon \to 0$, to a singular Hamiltonian with a potential which vanishes in the interior of the billiard domain $D$ and is strictly positive on its boundaries. In [19] we proved that under some natural conditions (they are satisfied by the potentials we encountered in the physics literature) the motion under the smooth Hamiltonian will smoothly limit, as $\varepsilon \to 0$, to the motion of the singular billiard system as long as one considers a *finite number of regular reflections* (reflections which are bounded away from the corners and from being tangent to the boundary). This result implies, in particular, that regular non-parabolic periodic orbits of the billiard are preserved, as their stability types do. Thus, if the billiard is dispersing (i.e. the billiard's boundary is composed of dispersing arcs intersecting at a non-zero angle), many unstable periodic orbits co-exist in the smooth Hamiltonian flow. However, under the same conditions, the phase space structure of the billiard flow and of its smooth Hamiltonian approximation may be of completely different character; we

*Date*: September 23, 2002.

proved in [19] that in an arbitrarily fine smooth approximation of any dispersing billiard, stability islands may be born from periodic trajectories which are *tangent to the billiards boundary* at some point. Furthermore, we conjectured that billiards with tangent periodic orbits are dense among dispersing billiards, and hence that the birth of stability islands in smooth approximations of dispersing billiards for arbitrarily small $\varepsilon$ is a typical phenomenon. Indeed, the billiards hyperbolicity implies [10, 1] that any dispersing billiard has many nearly tangent hyperbolic periodic orbits (of large period). Therefore, making them actually tangent to the boundary by slightly changing the shape of the boundary arc near an appropriately chosen point seems to be easy.

The appearance of elliptic islands in smooth Hamiltonians with steep repelling potentials may be counter-intuitive physically, yet it is not surprising from a mathematical point of view. Indeed, the billiard is a singular dynamical system, and the uniform hyperbolic structure of the dispersing billiard cannot survive a smoothening (softening) of the billiard; a neighborhood of the singularities is exactly the place where the elliptic islands emerge. Analogous results for the standard map were obtained in [4]. The possible appearance of elliptic islands in smooth approximations to two-dimensional billiards was suggested by numerical experiments in [8]. Their appearance in axially symmetric finite range potentials was analyzed in [2, 3]. In [19], the geometric mechanism for the creation of elliptic islands by tangent orbits (periodic and homoclinic) was suggested. In [14] this lead to a precise analysis, which included a sharp estimate on the island size (typically algebraic in the smoothing parameter $\varepsilon$) and scaling for arbitrary scattering billiard geometry and for physically relevant potentials.

However, there is another way for a hyperbolic billiard orbit to be destroyed by a singularity, namely when it falls into a corner of the billiard. The study of the effect of the corners on the behavior of the smooth Hamiltonians is the subject of the present paper.

In part, our work was inspired by recent experiments with soft billiards reported in [9]. In the experiments, a billiard domain is drawn by a fast moving laser beam which encloses very cold atoms. A small gap is opened after initial run time, and the decay rate of the remaining atoms supplies hints regarding the particles dynamics. By creating integrable vs. chaotic billiards and by varying the width of the laser beam one may examine the effects of chaotic motion and the effect of islands. Furthermore, in the numerical experiments which simulate the experiment, islands associated with tangent periodic orbits and islands associated with corner polygons are clearly observed. These experiments suggest that corner islands may be rather large.

We begin with a precise formulation of the work and with statements of the main results in a non-technical way.

## 2. Formulation and Main results:

2.0.1. *Billiard-like potentials:* Consider the 2-degrees-of-freedom Hamiltonian system defined by (1.1):

$$\ddot{x} = -\frac{\partial V(x,y;\varepsilon)}{\partial x}, \qquad \ddot{y} = -\frac{\partial V(x,y;\varepsilon)}{\partial y}, \tag{2.1}$$

where $V(x,y;\varepsilon)$ is a smooth ($C^{r+1}$) function of $(x,y)$ and $\varepsilon$. Consider the level set $H = h$. Let $D$ be a region in the $(x,y)$-plane with a piece-wise smooth boundary composed of $N$ smooth ($C^{r+2}$) arcs $S_1,\dots,S_N$. The points where two neighboring boundary arcs are joined are called the corner points. We assume that at all the corners the arcs meet at a non-zero angle less than $\pi$. Let $V(x,y;\varepsilon)$ limit to the billiard potential associated with $D$:

$$\lim_{\varepsilon\to+0} V(x,y;\varepsilon) = \begin{cases} 0 & \text{at} \quad (x,y)\in D, \\ c>h & \text{at} \quad (x,y)\in \partial D, \end{cases} \tag{2.2}$$

where $c$ may be infinite.

We assume that the singular behavior of the potential stems from its growth rate near the boundary alone, and not from its spatial structure, namely we assume that there exists a smooth *pattern function* $Q(x,y;\varepsilon)$ which has identical level sets to $V(x,y;\varepsilon)$ near each of the open arcs $S_1,\dots,S_N$ (excluding the corners) yet admits regular behavior (i.e. it has a finite smooth limit in a neighborhood of each of the open arcs $S_i$) in the limit $\varepsilon\to 0$. Then, for each $i=1,\dots,N$, there exists a *barrier function* $W_i(Q;\varepsilon)$ such that:

$$V(x,y;\varepsilon) = W_i(Q(x,y;\varepsilon);\varepsilon) \tag{2.3}$$

near each segment $S_i$. We also assume that the boundary arcs $S_i$ are level lines $\{Q(x,y;0)=0\}$, and we assume that for small $Q$

$$\nabla Q \neq 0. \tag{2.4}$$

Let the functions $Q$ be positive inside $D$, and assume that for small values of $Q$ the derivative $W'(Q)$ is bounded away from zero, uniformly for all small $\varepsilon$. Since $W$ must decrease as $Q$ increases across zero (see (2.2)), it follows that for small $Q$

$$W'(Q) < 0. \tag{2.5}$$

This means that we stick here to the case of the so-called *soft repulsion*, leaving the case of, say, Liennard-Jones potentials aside (or, equivalently, consider sufficiently large energies, far above the threshold energy for the existence of trapped orbits). Then, as it follows from (2.2), in any fixed energy level $\{H=h_0<h,\ \ h_0\neq 0\}$ the system under consideration *degenerates into the billiard in D as* $\varepsilon\to+0$.

Indeed, since the potential asymptotically vanishes inside $D$, on a finite distance from the boundary the motion becomes inertial as $\varepsilon\to+0$. When approaching the boundary the value of the potential sharply increases and the trajectory must be reflected. Furthermore, we have constructed our potential in such a way that its gradient ("the reaction force") is, asymptotically, normal to the boundary, which implies the standard reflection law ("the angle of incidence equals the angle of reflection"). Such kind of representation, in terms of pattern and barrier functions, was proposed for smooth billiard-approximating potentials in [19]. Precisely, we will adhere the following

**Definition 1.** *A family of $C^r$ potentials $V(x,y;\varepsilon)$ is called a billiard-like potential family if:*

- *There exists a domain D such that (2.2) is satisfied.*
- *There exist families of pattern functions $Q_i(x,y;\varepsilon)$ and of barrier functions $W_i(Q;\varepsilon)$ such that in an open neighborhood of the boundary of D without the corner point the following conditions are satisfied:*
  - *For sufficiently small $\varepsilon$ relations (2.3), (2.4), (2.5) hold.*
  - *As $\varepsilon\to 0$, the pattern function has a regular smooth limit in the $C^r$ topology.*
  - *As $\varepsilon\to 0$, for any finite, strictly positive values $V_1,V_2$, the functions $Q_i(W;\varepsilon)$ (defined as inverse to the barrier functions $W_i(Q;\varepsilon)$) tend to zero uniformly in the interval $W\in[V_1,V_2]$ along with all their $r+1$ derivatives.*

It was established in [19] that regular reflections of the billiard trajectories are regular limits, along with all the derivatives (up to the order $r$) with respect to the initial conditions, of trajectories of the Hamiltonians with the corresponding billiard-like potentials, whereas tangent segments of the billiard serve as limits of smooth trajectories in the $C^0$-topology. We will see that further conditions on the billiard-like potentials are needed so that a reasonable limiting flow near the corner will emerge.

2.0.2. *Main results:* Consider a billiard domain $D$ in which there exists a polygon which closes at a corner, and for which all other vertices correspond to regular billiard reflections from the billiard boundary. We call such a polygon a *corner polygon*, and denote it by $P_0$, see figure 1. Denote by $\theta$ the angle created by the billiard boundary arcs joining at the corner, and define $\phi_{in}, \phi_{out}$ as the angles created by the corner polygon with the corner bisector (notice the different direction of $\phi_{in}$ and $\phi_{out}$). The main question which we address here is under which conditions on $\phi_{in}, \phi_{out}, \theta$ and the potential the corner polygon will become a periodic orbit of the Hamiltonian flow (2.1) and, when it does, what is its stability. Notice that a segment connecting two different corners is equivalent to a polygon with two corner vertices, with $\phi_{in} = -\phi_{out}$ in each one of them. Here we deal with polygons going through one corner only.

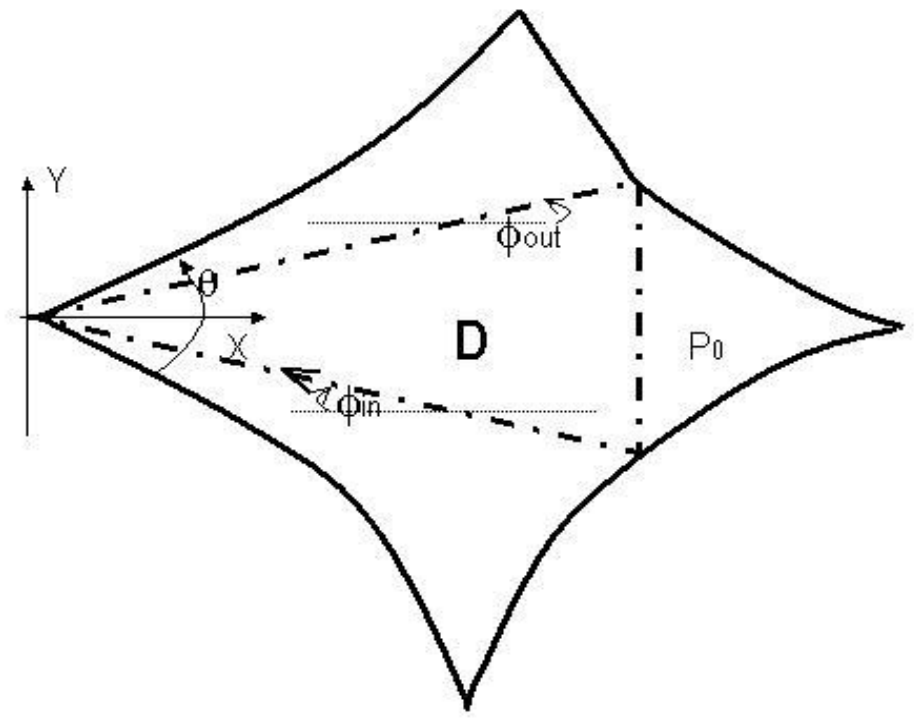

FIGURE 1. Geometry of a corner polygon.
$- - -$ is a corner polygon, $D$ is a dispersing billiard

In section 3 we describe the billiard motion near a corner. The computation shows that a billiard orbit which hits the boundary near the corner by the angle $\varphi$, exits a neighborhood of the corner after a finite number of reflections, and the angle which the outgoing trajectory makes with the corner bisector is close to one of two possible angles $\Phi_{\pm}(\varphi;\theta)$. The angle $\Phi_{+}(\varphi;\theta)$ is realized if the upper boundary is hit first, and $\Phi_{-}(\varphi;\theta)$ is realized otherwise.

In Theorem 1 of section 4 we prove that for any $\phi_{in}$ of the corner polygon, there is an interval $I$ such that if $\phi_{out} \in I$, then for sufficiently small $\varepsilon$ the Hamiltonian flow has a periodic orbit $P_\varepsilon$ which limits to $P_0$ as $\varepsilon \to 0$ (this requires an additional tuning of the pattern function $Q$, see details in Theorem 1). Moreover, $[\Phi_{-}(\phi_{in},\theta), \Phi_{+}(\phi_{in},\theta)] \subseteq I$, and we provide examples which show that strict inclusion is often possible. This fact is surprising. In particular, it shows that contrary to the previously studied cases (of non-singular periodic orbits and of tangent periodic orbits) *the existence of the periodic orbit which limits to a corner polygon is not determined by the billiard geometry alone*.

To describe the behavior of smooth billiard-like systems near the corners, we introduce an additional ingredient, the *scattering function.* This function captures the main features of the scattering by the potential at the corner point. To define the scattering function, we make some natural scaling assumption on the potential $V$ near the corner. Let $(x,y)$ denote Cartesian coordinates with the $x$-axis being the bisector of the billiard corner, and the origin at the corner point, see figure 1. We assume there exists a scaling

$$(\bar{x},\bar{y}) = \frac{1}{\delta(\varepsilon)}(x-x_\varepsilon, y-y_\varepsilon)$$

such that in the rescaled coordinates the potential has a finite limit as $\varepsilon \to 0$:

$$V(x_\varepsilon + \delta\bar{x}, y_\varepsilon + \delta\bar{y}; \varepsilon) \to V_0(\bar{x},\bar{y}).$$

Let the level set $V_0(\bar{x},\bar{y}) = h$ be a hyperbola-like curve, which asymptotes the lines $\bar{y} = \pm\bar{x}\tan\frac{\theta}{2} + c_\pm$ as $\bar{x} \to \infty$. This level curve bounds an open wedge $V_0 \leq h$ which extends towards $\bar{x} = +\infty$. For the rescaled system given by the Hamiltonian

$$H = \frac{1}{2}(p_x^2 + p_y^2) + V_0(\bar{x},\bar{y}), \tag{2.6}$$

every trajectory with the energy $H = h$ lies in this wedge.

Under some natural assumptions on $V$, we show that the solutions to the rescaled equations go towards $\bar{x} = +\infty$ as $t \to +\infty$ and as $t \to -\infty$, and that they always have an asymptotic incoming ($\varphi_{in} = -\lim_{t\to-\infty} \arctan\frac{p_y(t)}{p_x(t)}$, $|\varphi_{in}| \leq \frac{\theta}{2}$) and outgoing angles ($\varphi_{out} = \lim_{t\to+\infty} \arctan\frac{p_y(t)}{p_x(t)}$, $|\varphi_{out}| \leq \frac{\theta}{2}$). Moreover, there is a well defined limiting scattering function $\varphi_{out} = \Phi(\varphi_{in}, \eta)$ where $\eta$ is a scattering parameter of a parallel beam entering the wedge at $x = +\infty$ with incoming angle $\varphi_{in}$.

This scattering function $\Phi$ carries the needed information on the dynamics near the corner. For example, the range of $\Phi(\varphi_{in}, \cdot)$ is exactly the interval $I$ of allowed outgoing angles. So, according to Theorem 1, a billiard corner polygon with the ingoing angle $\phi_{in}$ and the outgoing angle $\phi_{out}$ may produce a periodic orbit of the Hamiltonian flow (2.1) at small nonzero $\varepsilon$ if and only if $\phi_{out} = \Phi(\phi_{in}, \eta)$ for some $\eta$. More precisely, given a $\phi_{out} \in I$ there exists a set (discrete, in general) of $\eta$'s such that $\phi_{out} = \Phi(\phi_{in}, \eta)$. Each of these values for which $\frac{\partial}{\partial\eta}\Phi(\phi_{in}, \eta) \neq 0$ corresponds to a limit of a family of hyperbolic periodic orbits $P_\varepsilon$ (provided the genericity condition of Theorem 2 is fulfilled).

If, on the contrary, $\phi_{out}$ corresponds to a maximum or minimum of $\Phi(\phi_{in}, \eta)$ as a function of $\eta$, then in a two-parameter family of Hamiltonians $H(x,y;\varepsilon,\gamma)$ ($\gamma$ is a parameter responsible for regular changes in the geometry of the billiard, i.e. it governs smooth changes in the pattern function $Q$ outside the corner points) there exists a wedge in the $(\gamma, \varepsilon)$-plane, at which the Hamiltonian flow possesses an elliptic periodic orbit which limits to the corner polygon as $\varepsilon \to +0$ (Theorem 3).

The stability of the corner-passing periodic orbits is solved here in terms of the scattering function $\Phi$ which is defined only by the potential at the corner, and it is almost independent of the geometrical properties of the underlying billiard (the above mentioned genericity condition is the only place where the geometry enters: this condition is always fulfilled if the billiard is dispersive and the corner polygon is never tangent to the boundary, while in the non-dispersive billiard where the boundary contains convex components this condition may be violated, but it may always be achieved by a small smooth perturbation of the boundary).

Unfortunately, there seem to be no explicit formulas which would relate the scattering function to the potential $V_0$. In particular, it is known [11] that in the case $V_0(\bar{x},\bar{y}) = e^{\bar{y}-k\bar{x}} + e^{-\bar{y}-k\bar{x}}$ (here $k = \tan\frac{\theta}{2}$, so $k \in (0,1)$) the system given by (2.6) has no other analytic integrals which are polynomial in momenta for $k \neq 1$ and $k \neq 1/\sqrt{3}$ (i.e. when the corner angle $\theta$ differs from $\pi/2$ and $2\pi/3$). The non-existence of meromorphic integrals for this system is proven in [21] (based on the method of [22]) for $k \neq 4/(m(m-1))^2, m \in Z$. Of course, it is straightforward to recover $\Phi$ numerically.

What we prove analytically (Lemma 1) is that $\Phi(\varphi,\eta)$ is a smooth function, and that as $\eta \to \pm\infty$ it approaches the billiard scattering angles $\Phi_\pm(\varphi;\theta)$. $\Phi$ can be shown to be non-monotone in quite natural examples. How to determine analytically the actual form of $\Phi$ and its critical values is, probably, an unsolvable question in the general case.

Finally, we find that there is one case in which we can prove the creation of elliptic islands by using only asymptotic information about the scattering function. This occurs when a billiard corner polygon bifurcates into a regular periodic orbit of the billiard: a billiard periodic orbit may detach from the corner point under a small perturbation of the boundary if and only if $\phi_{out} = \Phi_\pm(\phi_{in},\theta)$. In terms of the scattering function $\Phi$ this case corresponds to $\eta = \pm\infty$ and it is not covered by above mentioned Theorems 2 and 3. The behavior of the corner-passing periodic orbits of the Hamiltonian flow (2.1) at non-zero $\varepsilon$ has in this case a more profound relation with the billiard geometry. We analyze this problem and supply sufficient conditions for the creation of elliptic islands in the Hamiltonian flow in Theorem 4.

## 3. Billiard motion near corners.

Consider the billiard motion in an open angle (angle created by two rays). The usual representation of the billiard mapping by which the position and incidence angle serve as phase space variables is clearly ill-defined at the corner. Hence, we first introduce convenient variables. Let $(x,y)$ be Cartesian coordinates with the origin at the corner and with the $x$-axis along the bisector of the corner's angle $\theta$, directed into the billiard domain. Recall that we assume $\theta < \pi$. Let

$$k = \tan\frac{\theta}{2}.$$

Consider the billiard motion in the open angle $\{|y| \leq kx,\ x \geq 0\}$. Take a point $(x_0,y_0)$ within the angle and consider the billiard trajectory which starts at this point with the momenta $(p_x = -\sqrt{2h}\cos\varphi_{in}, p_y = \sqrt{2h}\sin\varphi_{in})$; we keep this choice of the directionality of $\varphi_{in}$ throughout the paper because it proves to be convenient when working with dispersive billiards (see the corollary to theorem 4). The following facts are well-known. The reader may easily recover their proofs by means of the following procedure: each time the billiard trajectory hits the boundary, let it not make a reflection but enter a copy of the angle obtained by the reflection of the angle, as a whole, with respect to this boundary. As a result, one gets a number of consecutive copies of the angle, intersected by a straight-line (instead of the polygonal trajectory in the single angle) and analysis of this picture is straightforward, see figure 2. Consider first the dependence of the outgoing direction on the initial conditions:

1. If $|p_y| \leq kp_x$ (i.e. $\pi - \frac{\theta}{2} < |\varphi_{in}| \leq \pi$), then the orbit never hits the boundary - it goes to infinity keeping the values of momenta constant.

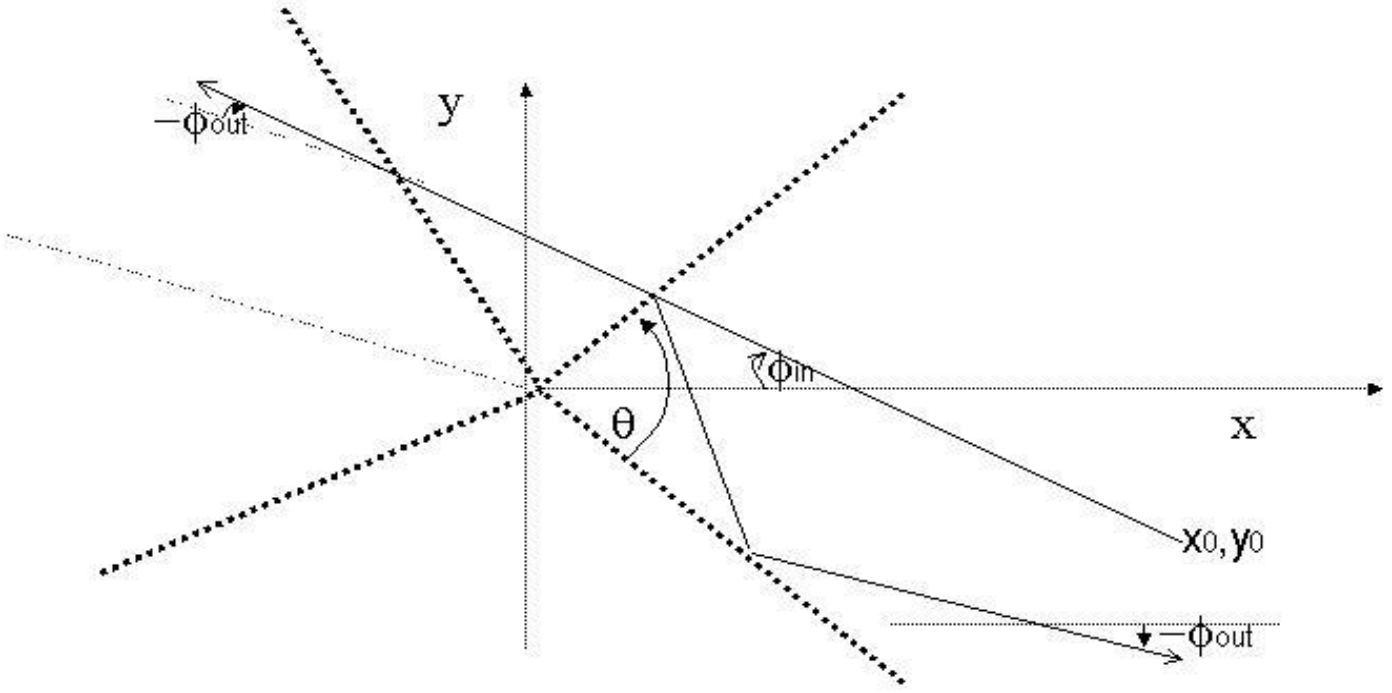

FIGURE 2. Billiard motion near an open angle.
$\cdots$ copies of the angle $\theta$, — billiard trajectory in the extended space and the resulting motion restricted to one angle.

2. If $-\arctan\frac{y_0}{x_0} < \varphi_{in} < \pi - \frac{\theta}{2}$, then the orbit hits the upper boundary first, makes $n_+ = \left]\frac{\pi-\varphi_{in}}{\theta} - \frac{1}{2}\right[$ reflections (we use the notation $]z[$ for the least integer which is not less than $z$), and then goes to infinity with the final values of momenta $(p_x = \sqrt{2h}\cos\varphi_{out}, p_y = \sqrt{2h}\sin\varphi_{out})$ where the outgoing angle is defined as $\varphi_{out} = \Phi_+(\varphi_{in}) = (-1)^{n_+}(\pi - \varphi_{in} - n_+\theta)$.

3. If $-\pi + \frac{\theta}{2} < \varphi_{in} < -\arctan\frac{y_0}{x_0}$, then the orbit hits the lower boundary first, makes $n_- = \left]\frac{\pi+\varphi_{in}}{\theta} - \frac{1}{2}\right[$ reflections, and goes then to infinity with the outgoing angle $\varphi_{out} = \Phi_-(\varphi_{in}) = (-1)^{n_-}(-\varphi_{in} - \pi + n_-\theta)$.

Summarizing, any orbit which does not enter the corner point (i.e. with $\varphi_{in} \neq -\arctan\frac{y_0}{x_0}$) goes out towards $x = +\infty$ after only a finite number ($n_\pm$) of reflections, and this number is bounded uniformly for all $(x_0, y_0, \varphi_{in})$ and $\theta$ provided $\theta$ is bounded away from zero. The final outgoing direction, called the *exiting direction*, is a uniquely defined function of $(x_0, y_0, \varphi_{in})$:

$$\Phi_\pm(\varphi_{in}) = \begin{cases} \pi - \varphi_{in} & \pi - \frac{\theta}{2} < |\varphi_{in}| \leq \pi \\ (-1)^{n_\pm}(-\varphi_{in} \pm \pi \mp n_\pm\theta) & -\pi + \frac{\theta}{2} < \varphi_{in} < \pi - \frac{\theta}{2} \end{cases}, \tag{3.1}$$
$$n_\pm = \left]\frac{\pi \mp \varphi_{in}}{\theta} - \frac{1}{2}\right[.$$

Let us pay a special attention to the range of ingoing angles $|\varphi_{in}| < \frac{\theta}{2}$ which corresponds to the orbits coming from infinity. Denote

$$N_\theta = \left[\frac{\pi}{\theta}\right], \quad X_\theta = \frac{\pi}{\theta} - \left[\frac{\pi}{\theta}\right], \tag{3.2}$$

and notice that

$$n_+(\varphi) = \begin{cases} N_\theta + 1 & \text{for} \quad -\frac{\theta}{2} < \varphi < \varphi_+^c \\ N_\theta & \text{for} \quad \varphi_+^c < \varphi < \frac{\theta}{2} \end{cases} \tag{3.3}$$

$$n_-(\varphi) = \begin{cases} N_\theta & \text{for} \quad -\frac{\theta}{2} < \varphi < \varphi_-^c \\ N_\theta + 1 & \text{for} \quad \varphi_-^c < \varphi < \frac{\theta}{2} \end{cases}$$

where

$$\varphi_\pm^c = \pm(X_\theta - \frac{1}{2})\theta.$$

Hence, depending on the numerical properties of $\frac{\pi}{\theta}$, four different types of corner angles emerge, corresponding to even/odd $N_\theta$ (indeed, $sign(\Phi_+(\varphi_{in}) - \Phi_-(\varphi_{in})) = (-1)^{N_\theta}$) and positive/negative values of $\left(\frac{1}{2} - X_\theta\right)$ (indeed, $sign(\varphi_+^c - \varphi_-^c) = sign(X_\theta - \frac{1}{2})$). The corresponding angles $\Phi_\pm(\varphi_{in})$ are shown in figure 3. We have thus established a complete understanding of the dependence of the exiting direction on initial conditions.

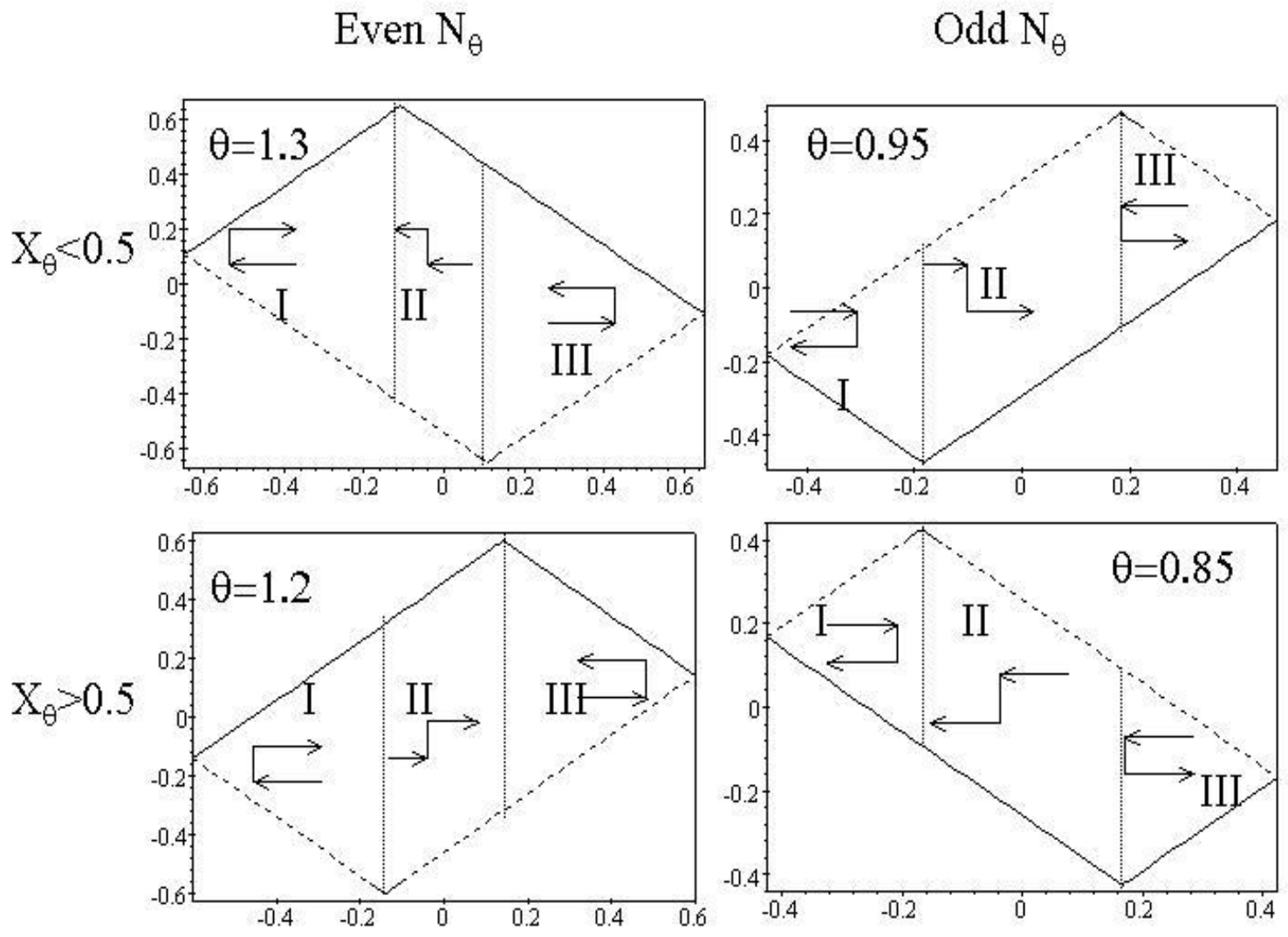

FIGURE 3. Reflections from an open angle.
Horizontal axis is $\varphi_{in}$, —– is $\Phi_+(\varphi_{in})$, - - - is $\Phi_-(\varphi_{in})$.
$\varphi_\pm^c$ are denoted by the dotted line and separate the different regions of $\varphi_{in}$ as listed in Table 1.

Now, fix a cross-section $x = R > 0$. The orbit whose all reflection points lie in the region $x < R$ will intersect the cross-section exactly in two points: $y = y_{in}$ and $y = y_{out}$. If $y_{in} < -R\tan\varphi_{in}$, then the lower boundary is hit first, and the upper boundary is hit first otherwise. It can be shown that the value of $y_{out}$ is given by the following formula (in particular, $y_{out}$ is piece-wise linear in $y_{in}$):

$$y_{out}^\pm = y_{in}\frac{\cos\varphi_{in}}{\cos\varphi_{out}}(-1)^{n_\pm+1} + R\left(\tan\varphi_{out} + (-1)^{n_\pm+1}\tan\varphi_{in}\frac{\cos\varphi_{in}}{\cos\varphi_{out}}\right). \tag{3.4}$$

Notice also that the distance from the orbit to the corner remains bounded from below by $K\sqrt{R^2+y_{in}^2}$ where the factor $K>0$ is bounded away from zero provided $\varphi_{in}+\arctan\frac{y_{in}}{R}$ is bounded away from zero.

Now we examine the action of the map $(y_{in},\varphi_{in})\mapsto(y_{out},\varphi_{out})$ on a parallel beam. On the cross-section $x=R$, this corresponds to the straight-line segment $\{\varphi_{in}=const,|y_{in}|\leq R\}$. Notice that for $|\varphi_{in}|<\frac{\theta}{2}$ the sign of the slope of $y^{\pm}_{out}(y_{in};\varphi_{in})$ (sign of $\frac{\partial y^{\pm}_{out}}{\partial y_{in}}$) has the same sign as the slope of $\Phi_{\pm}(\varphi_{in})$ (the signs of $\varphi_{in},\varphi_{out}$ are chosen to preserve this property). The corresponding graphs of the curves $(y_{out}(y,\varphi_{in}),\varphi_{out}(y,\varphi_{in}))$ are shown in table 1 and in figure 3, where arrows indicate the direction of increasing $y$ (in the table, the $y$-axis is horizontal and the $\varphi$-axis is vertical). The curves $(y_{out}(y,\varphi_{in}),\varphi_{out}(y,\varphi_{in}))$ are discontinuous, and, depending on the value of $\varphi_{in}$ (and the numerical properties of $\pi/\theta$), they either fold onto themselves or create a step as shown.

| | $N_\theta$ Even | | $N_\theta$ Odd | |
|---|---|---|---|---|
| $X_\theta<1/2$ | I. $\varphi_{in}<\varphi^c_+$ | 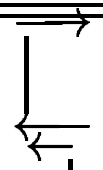 | I. $\varphi_{in}<\varphi^c_+$ | 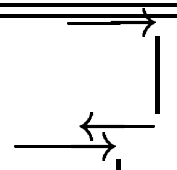 |
| | II. $\varphi^c_+<\varphi_{in}<\varphi^c_-$ | 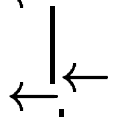 | II. $\varphi^c_+<\varphi_{in}<\varphi^c_-$ | 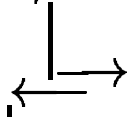 |
| | III. $\varphi^c_-<\varphi_{in}$ | 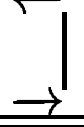 | III. $\varphi^c_-<\varphi_{in}$ | 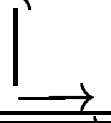 |
| $X_\theta>1/2$ | I. $\varphi_{in}<\varphi^c_-$ | 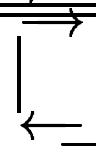 | I. $\varphi_{in}<\varphi^c_-$ | 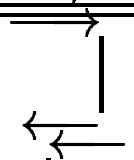 |
| | II. $\varphi^c_-<\varphi_{in}<\varphi^c_+$ | 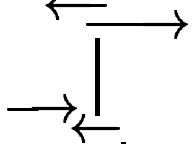 | II. $\varphi^c_-<\varphi_{in}<\varphi^c_+$ | 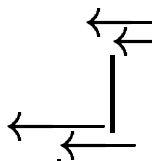 |
| | III. $\varphi^c_+<\varphi_{in}$ | 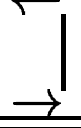 | III. $\varphi^c_+<\varphi_{in}$ | 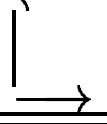 |

By now, we have defined *the corner map* $T^0_{cor}:(y_{in},\varphi_{in})\mapsto(y_{out},\varphi_{out})$ for the billiard in the open angle. Analogously, one can define the corner map near the corner point of any billiard, with a curvilinear boundary. We just take $R$ sufficiently small, then for the orbits which hit the boundary at $x<R$ the effect of curvature will be only a small (of order $x$) additional rotation of the vector of momenta plus a small ($o(x)$) displacement in $(x,y)$ at each reflection. Since the number of reflections is finite, it follows that near the corner the orbits of a curvilinear and the corresponding linear billiards remain close, provided $\varphi_{in}$ and $\varphi_{out}$ are bounded away from $\pm\frac{\theta}{2}$. Therefore, for small $R$, the map $T^0_{cor}$ is defined for the curvilinear billiard as well, and the relation between $(y_{in},\varphi_{in})$ and $\varphi_{out}$ will be $O(R)$-close to that given by (3.1), while $y_{out}$ will be $o(R)$ close to that given by (3.4) (at least for orbits which are nonparallel to the boundary). The effect of the curvature on the corner polygon for small deviations of $(y_{in},\varphi_{in})$ in the small $R$ limit may be explicitly calculated. Let $\kappa_\pm$ denote the curvature on the upper and lower boundaries of the corner. We choose the sign of $\kappa$ in such a way that $\kappa>0$ for the concave boundary arcs (when looked from within the billiard domain). Then, it may be shown that

$$\lim_{R\to0}DT^0_{cor}\begin{pmatrix}\Delta\varphi\\ \Delta y\end{pmatrix}=(-1)^{n_\pm+1}\begin{pmatrix}1 & 2\cos\phi_{in}\sum\frac{\kappa_{(-1)^{j+1}}}{\cos\alpha_j}\\ 0 & \frac{\cos\phi_{in}}{\cos\phi_{out}}\end{pmatrix}\begin{pmatrix}\Delta\varphi\\ \Delta y\end{pmatrix} \tag{3.5}$$

where

$$\alpha_j=\frac{\pi}{2}-\frac{\theta}{2}-j\theta-\phi_{in}. \tag{3.6}$$

For the dispersive curvilinear billiard, when $\kappa > 0$ for all boundary arcs, the image of the parallel beam is always divergent [16, 20], i.e. every continuous piece of the curve $(y_{out}(y_{in}), \varphi_{out}(y_{in}))$ must be a graph of a strictly increasing function of $\varphi_{out}$ vs. $y_{out}$. At the same time, as we explained above, this curve must be close to that we obtain in the straight-linear case as shown schematically for one case ($N_\theta$ even, $\varphi_{in} > \varphi^c_\pm$) in figure 4.

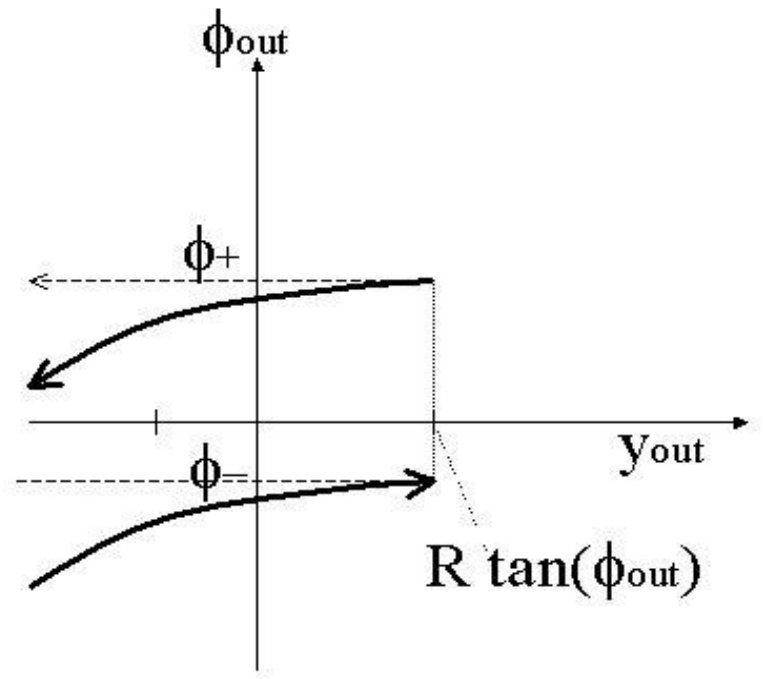

FIGURE 4. Reflection of a parallel beam from a corner.
- - reflection from an open angle corner, —— reflection from a corner created by dispersing arcs.
arrow indicates the direction of increase of $y_{in}$.

Consider a Hamiltonian flow which gives a sufficiently good approximation to the billiard flow away from a small neighborhood of the corner. Then, $\{x = R\}$ serves as a cross-section for the Hamiltonian flow as well, and the corresponding Poincaré map $T^\varepsilon_{cor}$ is close to the billiard corner map $T^0_{cor}$ away from a small neighborhood of the discontinuity line $\varphi_{in} = -\arctan\frac{y_{in}}{R}$. The image of the horizontal line (the parallel beam) by the Poincaré map is a continuous line which approximates the image of the same line by the map $T^0_{cor}$. Examining table 1 and the corresponding figures like 4, we see that in the dispersing case the image of this line is non-monotone in $y_{in}$ in many cases (when $X < 0.5$ for all entering angles and when $X > 0.5$, when $\varphi_{in} \notin [\varphi^c_+, \varphi^c_-]$).

This non-monotonicity suggests that the assumption in Theorem 3 of the occurrence of the extrema in the scattering function is natural. Furthermore, the non-monotonicity implies that the map $T^\varepsilon_{cor}$ creates a horseshoe-like shape. More precisely, we always have an interval of values of $\varphi_{in}$ where the map $T^\varepsilon_{cor}$ creates a fold in the parallel beam: for each $\varphi_{in}$ in this interval there exists $y^*(\varphi_{in})$ (and the corresponding $\varphi^*(\varphi_{in}) = \varphi^\varepsilon_{out}(\varphi_{in}, y^*(\varphi_{in}))$ such that $\left.\frac{\partial \varphi^\varepsilon_{out}}{\partial y}\right|_{(\varphi_{in}, y^*(\varphi_{in}))} = 0$. If the underlying billiard is dispersive, then by transitivity one can expect that the orbit exiting the corner with $\varphi = \varphi^*(\varphi_{in})$ will return back close to $\varphi = \varphi_{in}$ after a number of regular reflections. Using the cone-preservation property (see [16, 20, 19]) of the billiard flow for dispersive billiards, one can show that the fold in the image of the parallel beam is preserved after any number of regular reflections. For sufficiently small $\varepsilon$ the same must be true for the smooth flow defined by the corresponding billiard-like potentials. Hence, we can expect a Smale horseshoe here and, in particular, the birth of elliptic periodic orbits like in Theorem 3. Exact formulation of some of these ideas (for the case in which the corner polygon satisfies $\phi_{out} = \Phi_{\pm}(\phi_{in})$, so no ergodicity arguments are needed) is given in Theorem 4.

## 4. Existence of a periodic orbit near a billiard corner polygon

Consider a billiard-like Hamiltonian system (2.1) which degenerates at $\varepsilon = 0$ into a billiard in a domain $D$. Take some corner point and let $P_0$ be a corner polygon: a polygon which leaves the corner with some outgoing angle $\phi_{out}$, makes a finite number of *regular* billiard reflections and then closes at the corner, entering it with the ingoing angle $\phi_{in}$.

Let us choose some small $R$ and consider the cross-section $x = R$. The orbit $P_0$ intersects it at two points: $y_{out} = R\tan\phi_{out}$ and $y_{in} = -R\tan\phi_{in}$. The billiard flow in the region $x > R$ defines an *external billiard map* $T^0_{ext}$ which acts on the phase plane corresponding to the initial conditions on the cross-section and maps a small neighborhood of the point $(y_{out}, \phi_{out})$ into a small neighborhood of the point $(y_{in}, \phi_{in})$, as shown in figure 5. Since we assume that $P_0$ is a non-tangent orbit, this map is locally smooth, and, moreover, it depends smoothly on the shape of the billiard domain.

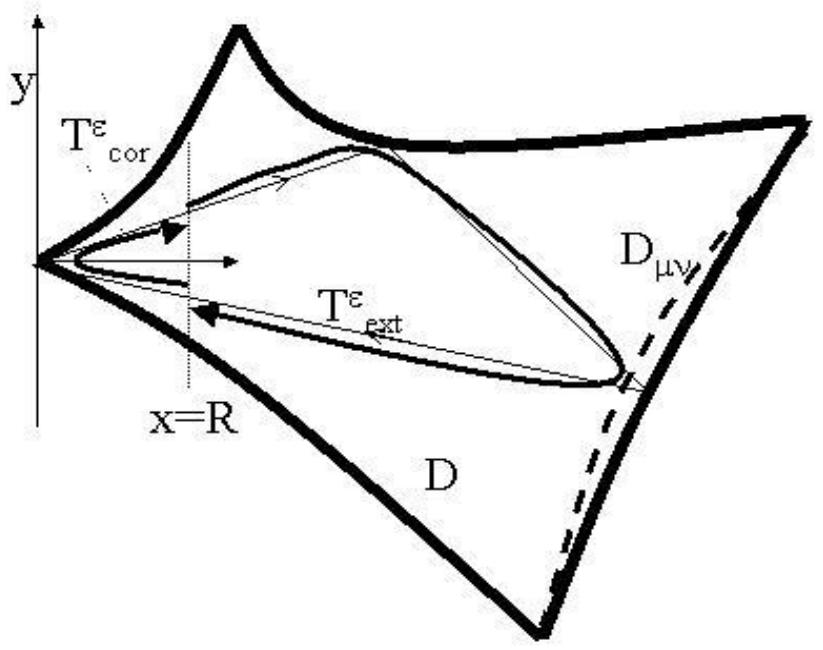

FIGURE 5. Definition of external and corner maps.

Include the billiard domain $D$ in a two-parameter family of domains $D_{\mu\nu}$, by including the pattern function $Q(x,y,\varepsilon)$ in a smooth two-parameter family of functions $Q(x,y,\varepsilon;\mu,\nu)$; the boundary of $D_{\mu\nu}$ is given by zero level lines of $Q(x,y,0;\mu,\nu)$ (see for example figure 5). Assume that all the functions $Q(\cdot;\mu,\nu)$ coincide in a small neighborhood of the corners, while outside the small neighborhood of the corners the dependence on $\mu$ and $\nu$ is *generic* so that

$$\frac{\partial T^0_{ext}(y_{out},\phi_{out})}{\partial(\mu,\nu)} \neq 0. \tag{4.1}$$

This condition is insensitive to the precise choice of the small $R$ (i.e. to the precise position of the cross-section). The corresponding families of potentials $V(\cdot,\varepsilon;\mu,\nu)$ thus constructed will be called embedding billiard-like families for $V(x,y;\varepsilon)$.

**Definition 2.** *A corner orbit* **produces a periodic orbit** *if any family of embedding billiard-like potentials $V(\cdot,\varepsilon;\mu,\nu)$, has a continuous (in $\varepsilon$) family of potentials $V(\cdot,\varepsilon,\mu(\varepsilon),\nu(\varepsilon))$ such that for all small $\varepsilon>0$ the corresponding flow has a periodic orbit $P_\varepsilon$ such that $P_\varepsilon\to P_0$ as $\varepsilon\to+0$.*

**Definition 3.** *A corner point is* **non-sticky** *if there exists a small neighborhood of it such that for all small $\varepsilon>0$, any trajectory of the Hamiltonian system which enters this neighborhood exits it in a finite time.*

A sufficient condition for the corner to be non-sticky is that $V_x'(x,y;\varepsilon)<0$ for all small $x$ (recall that we put the origin at the corner).

**Theorem 1.** *Consider a billiard-like Hamiltonian system (2.1) with non-sticky corners. Then, for every $\phi_{in}$ there exists an interval $I(\phi_{in})$ such that a corner orbit produces a periodic orbit if and only if $\phi_{out}\in I(\phi_{in})$. Furthermore, $[\Phi_{-(-1)^{N_\theta}}(\phi_{in}),\Phi_{+(-1)^{N_\theta}}(\phi_{in})]\subseteq I(\phi_{in})$.*

*Proof.* Consider two small cross-sections $\Sigma^\pm$ to the corner orbit in the phase space, $\Sigma^+$ intersects the outgoing segment of the orbit and $\Sigma^-$ intersects the ingoing one, both the cross-sections lying in $\{x=R\}$. The phase space is parametrized by the position $(x,y)$ of the point and its momenta, and fixing the energy level $H=h$ the values of the ingoing (outgoing) momenta are uniquely restored from the angle $\varphi$ of $\Sigma^-$ (respectively, $\Sigma^+$) which defines the direction of motion: $p_x|_{\Sigma^\pm}=\pm\sqrt{2(h-V(x,y))}\cos\varphi$, $p_y|_{\Sigma^\pm}=\sqrt{2(h-V(x,y))}\sin\varphi$. Compute the return map from $\Sigma^-$ to itself near $(\varphi=\phi_{in}, y=-R\tan\phi_{in})$ in two steps. First construct the corner map

$$T^\varepsilon_{cor}:\Sigma^-\to\Sigma^+$$
$$(\varphi,y)\to(\varphi_{cor}=F(\varphi,y;\varepsilon),y_{cor}=G(\varphi,y;\varepsilon)).$$

Since the corner is non-sticky, every trajectory starting with $x=R$ towards the corner must return to the cross-section after a finite time. Therefore, the map $T^\varepsilon_{cor}$ is well defined and $C^r$ for $\varepsilon>0$.

Take any $\varphi(\varepsilon)\to\phi_{in}$ and any $y(\varepsilon)$ such that $F(\varphi(\varepsilon),y(\varepsilon);\varepsilon)$ has a limit as $\varepsilon\to+0$. Denote the set of all such limiting points by $J(\phi_{in},R)$. By continuity of $F$ it follows that $J(\phi_{in},R)$ is a closed interval.

We now prove that $\Phi_-(\phi_{in})$ and $\Phi_+(\phi_{in})$ belong to $J(\varphi_{in},R)$ for any $R$. Indeed, let $(\varphi(\varepsilon),y(\varepsilon))=(\phi_{in},-R\tan\phi_{in}+y_0)$, where $y_0$ is non-zero and small (independent of $\varepsilon$). Then, as it was explained in the previous section, the billiard trajectory starting with these initial conditions will stay away (at a distance of order $|y_0|$ at least) from the corner and the number of reflections before returning to the cross-section will be finite and all the reflections will be non-tangent (i.e. at non-zero angles). Hence, according to [19], the corresponding trajectory of the Hamiltonian flow tends, as $\varepsilon\to+0$, to the billiard trajectory. Therefore, the corresponding value of $\varphi_{cor}$ must be close to $\Phi_{\mathrm{sign}(y_0)}(\phi_{in})$ (see (3.1)), and $\varphi_{cor}$ will indeed approach $\Phi_{\mathrm{sign}(y_0)}(\phi_{in})$ as $y_0\to0$. By continuity, all intermediate values between $\Phi_-(\phi_{in})$ and $\Phi_+(\phi_{in})$ lie in $J(\phi_{in},R)$ as well.

Define $I(\phi_{in})=\cap_{0<R<R_0}J(\phi_{in},R)$. As we proved above, $[\Phi_-(\phi_{in}),\Phi_+(\phi_{in})]\subseteq I(\phi_{in})$. Furthermore, by construction of $J(\phi_{in},R)$, if there exists a family of Hamiltonians with trajectories which limit to the corner orbit, then the corner orbit must satisfy $\phi_{out}\in I(\phi_{in})$.

Now construct the map $T^{\varepsilon}_{ext} : \Sigma^{+} \to \Sigma^{-}$ by the Hamiltonian flow near the external part of the corner orbit (i.e. the part which lies outside a neighborhood of the corner). According to [19], since all the reflections are non-tangent, the map $T^{\varepsilon}_{ext}$ is $C^r$-smooth and close, in the $C^r$-topology, to the corresponding billiard map $T^{0}_{ext}$. Therefore the map $T^{\varepsilon}_{ext}$ may be written in the form (recall that $\phi_{in}, \phi_{out}, y_{in}, y_{out}$ are determined by the corner orbit and are fixed):

$$T^{\varepsilon}_{ext} : \begin{pmatrix} \bar{\varphi} \\ \bar{y} \end{pmatrix} = T^{\varepsilon}_{ext}(\phi_{out}, y_{out}) + (T^{\varepsilon}_{ext})'(\phi_{out}, y_{out}) \cdot \begin{pmatrix} \varphi_{cor} - \phi_{out} \\ y_{cor} - y_{out} \end{pmatrix} + \dots \tag{4.2}$$

where the dots stand for the quadratic and higher order terms in $(\varphi_{cor} - \phi_{out}, y_{cor} - y_{out})$. Recall that we consider a two-parameter family of billiard domains, and since $T^{\varepsilon}_{ext}(\phi_{out}, y_{out})$ is close to $T^{0}_{ext}(\phi_{out}, y_{out})$, the genericity assumption (4.1) allows to assume that the parameters $(\mu, \nu)$ are chosen in such a way that

$$T^{\varepsilon}_{ext}(\phi_{out}, y_{out}) = \begin{pmatrix} \phi_{in} + \nu \\ y_{in} + \mu \end{pmatrix}.$$

Now, composing the external map $T^{\varepsilon}_{ext}$ and the corner map $T^{\varepsilon}_{cor}$ we obtain the following fixed point equation for the composed map:

$$\begin{pmatrix} \varphi \\ y \end{pmatrix} = T^{\varepsilon}_{ext} \circ T^{\varepsilon}_{cor} \begin{pmatrix} \varphi \\ y \end{pmatrix} = (T^{\varepsilon}_{ext})'(\phi_{out}, y_{out}) \cdot \begin{pmatrix} F(\varphi, y; \varepsilon) - \phi_{out} \\ G(\varphi, y; \varepsilon) - y_{out} \end{pmatrix} + \begin{pmatrix} \phi_{in} + \nu \\ y_{in} + \mu \end{pmatrix} + \dots$$

Choosing any $(\varphi, y) \in \Sigma^{-}$ the above equation defines $\nu, \mu$ for which these values correspond to a fixed point, which corresponds to a periodic orbit of the Hamiltonian flow by construction. If $\phi_{out} \in I(\phi_{in})$, we may choose the coordinates of the fixed point $(\varphi(\varepsilon), y(\varepsilon))$ such that $\varphi(\varepsilon) \to \phi_{in}$ and $F(\varphi(\varepsilon), y(\varepsilon); \varepsilon) \to \phi_{out}$. This would give $\nu(\varepsilon) \to 0$. Since $y = O(R)$, choosing $R = R(\varepsilon)$ tending to zero sufficiently slowly so that the above representation for the composed Poincaré map $T^{\varepsilon}_{ext} \circ T^{\varepsilon}_{cor}$ remains valid, we also ensure $\mu(\varepsilon) \to 0$. By construction, the periodic orbit which corresponds to such chosen values of $(\varphi(\varepsilon), y(\varepsilon))$ limits to the corner orbit as $\varepsilon \to +0$, hence the corner orbit indeed produces a periodic orbit. □

Notice that there are examples where the inclusion $[\Phi_{-(-1)^{N_\theta}}(\phi_{in}), \Phi_{+(-1)^{N_\theta}}(\phi_{in})] \subset I(\varphi)$ is strict, see section 5.3.1.

## 5. Local Analysis near corners.

Theorem 1 demonstrates that periodic orbits which are close to a billiard corner orbit are expected to appear in the smooth approximation to billiards if the incoming and outgoing directions at the corner are within some range. To obtain more precise information on the existence and stability of these periodic orbits in a given potential family the motion near the corners must be analyzed.

5.1. **The corner scaling assumption.** To understand the smooth motion near the corners, and in particular the nature of the corner mapping $T^{\varepsilon}_{cor}$, we need to understand the structure of our Hamiltonian system at the corner point. To this aim we rescale the equations of motion. The conditions on the potential by which this scaling simplifies the equations are summarized in the following *Corner Scaling condition.* Take a small $\delta$ and let

$$x = \delta\bar{x} + x_{\varepsilon}, \quad y = \delta\bar{y} + y_{\varepsilon}, \quad p_x = \sqrt{h}\,\overline{p_x}, \quad p_y = \sqrt{h}\,\overline{p_y}, \quad t = \frac{\delta}{\sqrt{h}}\bar{t}, \tag{5.1}$$

and

$$V_\varepsilon(\bar{x},\bar{y};h) = \frac{1}{h} V(\delta\bar{x} + x_\varepsilon, \delta\bar{y} + y_\varepsilon; \varepsilon)$$

The scaled Hamiltonian $\overline{H} = \frac{1}{h}H$ is

$$\overline{H} = \frac{1}{2}(\bar{p}_x^2 + \bar{p}_y^2) + V_\varepsilon(\bar{x},\bar{y};h), \tag{5.2}$$

and we consider the level set $\overline{H} = 1$ which in the configuration space corresponds to the region $V_\varepsilon(\bar{x},\bar{y},h) \le 1$[1]. Away from the corners this region is bounded by a level set of the pattern functions $Q_\varepsilon(\delta\bar{x} + x_\varepsilon, \delta\bar{y} + y_\varepsilon)$.

Take some sufficiently small $R > 0$ and consider the region

$$C_\varepsilon = \{(x,y) | x < R, V(x,y;\varepsilon) \le h\}.$$

Part of our assumptions on $V$ is that in the rescaled coordinates this region limits, as $\bar{x} \to +\infty, \varepsilon \to +0$, to a wedge with a limiting angle $\theta$, as in the billiard geometry. Namely, in the rescaled coordinates the region $C_\varepsilon$ is written as

$$\overline{C}_\varepsilon = \{(\bar{x},\bar{y}) | \bar{x} < (R - x_\varepsilon)/\delta, V_\varepsilon(\bar{x},\bar{y}) \le 1\}. \tag{5.3}$$

Define

$$\overline{C} = \lim_{\varepsilon_0 \to +0} \cap_{\varepsilon < \varepsilon_0} \overline{C}_\varepsilon.$$

**Condition 1.** *Assume there exists a function $V_0(\bar{x},\bar{y})$ defined in the region $\overline{C}$ such that for some functions $\delta(\varepsilon), x_\varepsilon, y_\varepsilon$ tending to zero as $\varepsilon \to 0$ the rescaled potential $V_\varepsilon(\bar{x},\bar{y})$ tends to $V_0$ as $\varepsilon \to 0$, uniformly along with all derivatives on any compact subset of $\overline{C}$. Furthermore, assume that for sufficiently large $\bar{x}$ the potential $V_\varepsilon(\bar{x},\bar{y})$ is of the form (recall that $k = \tan\frac{\theta}{2}$):*

$$V_\varepsilon(\bar{x},\bar{y}) = W_+(k\bar{x} - \bar{y}, \varepsilon) + W_-(k\bar{x} + \bar{y}, \varepsilon) + W_\varepsilon(\bar{x},\bar{y}) \tag{5.4}$$

*where $W_\varepsilon(\bar{x},\bar{y}) \to 0$ as $\bar{x} \to \infty$, and $W_\pm(r) \to 0$ as $r \to \infty$, uniformly and along with all derivatives (see (5.5)) for all sufficiently small $\varepsilon \ge 0$. Furthermore, $W'_\pm(r) < 0$, and there exist $\alpha > 0$ and $K > 0$ such that*

$$|\partial^n_{x^m, y^{n-m}} V_\varepsilon(\bar{x},\bar{y})| \le K(|k\bar{x} + \bar{y}|^{-(n+\alpha)} + |k\bar{x} - \bar{y}|^{-(n+\alpha)}) \tag{5.5}$$

*(here $m = 0, \dots, n$ and $n = 0, \dots, r+1$).*

Notice that the rescaled system is well defined and smooth at $\varepsilon = 0$. It is also seen that under this assumption the boundary of the region $\overline{C}$ (it corresponds to $V_0 = 1$) is given, as $x \to +\infty$, by two curves which approach asymptotically the lines $\bar{y} = k\bar{x} - c_+$ and $\bar{y} = -k\bar{x} + c_-$ where $c_\pm = W_\pm^{-1}(1)$ at $\varepsilon = 0$.

Let us take some sufficiently large positive $M$ and cut the wedge $\overline{C}$ by the line $\{\bar{x} = M\}$. We assume that $V_0$ is a *scattering* potential which means that any orbit starting at $\bar{x} \le M$ inside $\overline{C}$ with the energy $\overline{H} = 1$ leaves the region $\bar{x} \le M$ in a finite time.

We show below that if $M$ is sufficiently large, then the above assumption guarantees that every trajectory starting in $\overline{C}$ with $\overline{H} = 1$ tends to $\bar{x} = +\infty$ as $t \to \pm\infty$, i.e. the rescaled system at $\varepsilon = 0$ is indeed a scattering system. The non-stickiness of the corner point (see Theorem 1) also follows from this assumption. A sufficient condition is, of course, $\frac{\partial}{\partial \bar{x}} V_0 > 0$ everywhere in $\overline{C}$.

[1]Hereafter we will not show the explicit dependence of $V_\varepsilon(\bar{x},\bar{y},h)$ on $h$. In some cases one may choose a rescaling so that $V_\varepsilon$ is independent of $h$. Otherwise, the analysis and results apply to the range of $h$ values for which the assumptions on $V_\varepsilon$ hold.

Note that the corner scaling is different from the near tangency scaling that was used in [14], so these two scaling assumptions should be verified independently for near-tangent and near-corner trajectories respectively.

For example, take $V=\sum_{i=1}^{n}W(Q_i)$, where the level sets of neighboring arcs $(Q_i(x,y;\varepsilon)=0)$ intersect at non-zero angles (near the boundaries and away from the corners, we may write any billiard-like potential in this form). Then (5.5) is satisfied if $W(Q)=O(Q^{-\alpha})$ for some $\alpha>0$. While the work here applies to many natural physical potentials - e.g. $W=\varepsilon Q^{-\alpha}$ and $W=e^{-Q^{\beta}/\varepsilon}$, it excludes the case $W=\varepsilon|\ln Q|$ which was allowed in [19, 14] for the near-tangent orbits.

5.2. **Dynamics in the scaled equations of motion.** We first establish the asymptotic properties of the scaled Hamiltonian flow (5.2), establishing the existence of a scattering function which asymptotes to the billiard scattering functions $\Phi_{\pm}(\varphi_{in};\theta)$ in the appropriate limit. We then compute the corner map $T^{\varepsilon}_{cor}$ for the non-rescaled system (2.1). In this and in the next subsection, we drop all the bars from the rescaled variables. We start with the analysis of the behavior of the rescaled Hamiltonian flow (5.2) at large $x$.

**Proposition 1.** *Consider a family of billiard-like potentials satisfying the corner scaling assumption with a scattering rescaled potential $V_0$. For any initial condition $(x(0),y(0),p_x(0),p_y(0))$ of the rescaled equations with $(x(0),y(0))\in\overline{C}$ and the rescaled energy $H=1$, we have $x(t)\to\infty$ as $t\to\pm\infty,\varepsilon\to+0$ and the asymptotic incoming and outgoing angles:*

$$\tan\varphi_{in}=-\frac{p_y(-\infty)}{p_x(-\infty)}\qquad \tan\varphi_{out}=\frac{p_y(+\infty)}{p_x(+\infty)}$$

*are well defined and depend continuously on initial conditions.*

*Proof.* Here we only outline the main ideas, see appendix for the complete proof. For some large enough $M$, according to the scattering assumption, any trajectory of the rescaled system starting at $x\leq M$ will leave this region in a finite time (we made this assumption at $\varepsilon=0$ and, with $M$ fixed, it holds true for all small $\varepsilon$ due to the continuity in $\varepsilon$). Since the time of exit from the region $\{x\leq M\}$ is finite, the coordinates and momenta at the moment of exit depend continuously on the initial conditions and $\varepsilon$. So, it remains to prove the proposition for large enough initial values of $x$ and positive initial values of $p_x$ (this corresponds to the limit $t\to+\infty$, the limit $t\to-\infty$ is considered in an analogous way). To this aim, the wedge region $\overline{C}$ is divided to its bulk and to boundary layers of thickness $L$ which reside along the corner rays, and start at $x>M$. In Lemma 6 (see appendix) it is proved that outside of these boundary layers the momenta are preserved to order $O(L^{-\alpha/2})$. Hence, once we have proven (lemmas 7 and 8) that the distance $L(t)$ to the boundaries of $\overline{C}$ tends to infinity as $t\to+\infty$, we immediately obtain that the momenta must indeed stabilize in this limit. □

Note that an analogous statement can be found in [6, 7]; in essence, our rescaled potential $V_0$ is, at sufficiently large $x$, a small perturbation of the potential $W_{+}(kx-y)+W_{-}(kx+y)$, and the latter is a potential of the kinds considered in [7].

The following lemma proves the smoothness of the asymptotic angles. Note that a close result was obtained in [13] for a smaller class of potentials yet for any number of degrees of freedom and by a method which looks completely different from ours.

**Lemma 1.** $\varphi_{in}$ *(resp. $\varphi_{out}$) depends smoothly on the initial conditions provided* $|\varphi_{in}|<\frac{\theta}{2}$ *(resp. $|\varphi_{out}|<\frac{\theta}{2}$).*

*Proof.* We will prove this claim for $\varphi_{out}$ (the behavior of $\varphi_{in}$ is studied absolutely analogously). By proposition 1, any trajectory will achieve, at some time $t_0$, some sufficiently large value of $x$ and momenta values which are close to the limiting ones. Moreover, the values of the momenta will be almost preserved at all times larger than $t_0$. In particular, we have $p_x(t) > 0$ and $|p_y(t)| < kp_x(t)$ for $t \geq t_0$. It follows then that the distance to both boundaries grows with a non-zero velocity at $t \geq t_0$. Hence, by taking a larger value of $t_0$, if necessary, we may achieve that both the values $kx(t_0) \pm y(t_0)$ are sufficiently large. The values of $x(t_0)$, $y(t_0)$, $p(t_0)$ depend smoothly, of course, on the initial conditions. So we may assume that our orbit starts at $t = t_0$ with the initial values $x(t_0)$, $y(t_0)$, $p(t_0)$ and we will prove that $\varphi_{out}$ depends smoothly on these initial data.

Let us define the following boundary value problem. Given a time interval $[t_0, t_1]$, fix $x(t_0), y(t_0) \in \overline{C}$ with sufficiently large $x(t_0)$ and with $p(t_1) = (p_x(t_1), p_y(t_1))$ such that $|p_y(t_1)| < kp_x(t_1)$. We will prove that these data define the trajectory uniquely, for any $t_1 \geq t_0$ such that $(x(t), y(t))$ lie in the region $\overline{C}_\varepsilon$ (where the rescaled system is defined) at all $t \in [t_0, t_1]$; this includes the case $t_1 = \infty$ at $\varepsilon = 0$.

Indeed, rewrite equations (5.2) in the following form:

$$
\begin{aligned}
(5.6)\qquad x(t) &= x(t_0) + \int_{t_0}^{t} p_x(s)ds, \quad p_x(t) = p_x(t_1) + \int_{t}^{t_1} \frac{\partial V_\varepsilon(x(s), y(s))}{\partial x} ds,\\
y(t) &= y(t_0) + \int_{t_0}^{t} p_y(s)ds, \quad p_y(t) = p_y(t_1) + \int_{t}^{t_1} \frac{\partial V_\varepsilon(x(s), y(s))}{\partial y} ds.
\end{aligned}
$$

Define an operator $S : p(t) \mapsto \hat{p}(t)$:

$$
\begin{aligned}
x(t) &= x(t_0) + \int_{t_0}^{t} p_x(s)ds, \qquad y(t) = y(t_0) + \int_{t_0}^{t} p_y(s)ds,\\
(5.7)\qquad \hat{p}(t) &= p(t_1) + \int_{t}^{t_1} \nabla V_\varepsilon(x(s), y(s))ds.
\end{aligned}
$$

This operator acts on the space $U_\delta$ of continuous functions $p(t)$ defined at $t \in [t_0, t_1]$ and such that $\|p(t) - p(t_1)\| \leq \delta$ for some small $\delta$.

**Claim 1.** *If $|p_y(t_1)| < kp_x(t_1)$, then at sufficiently large $x(t_0)$ the operator $S$ takes the space $U_\delta$ into itself, and it is smooth and contracting on $U_\delta$, uniformly for all $t_0 \leq t_1 \leq +\infty$.*

*Proof.* If $p \in U_\delta$, then for sufficiently small $\delta$ we have $k\dot{x} \pm \dot{y}$ bounded away from zero for all $t \in [t_0, t_1]$. Hence we may use (5.5) to estimate the integrals in the momentum equations of (5.7):

$$
\begin{aligned}
\left| \int_{t}^{t_1} \nabla V_\varepsilon(x(s), y(s))ds \right| &< K \int_{t_0}^{t_1} \left( \left| (kx(s) + y(s))^{-(1+\alpha)} \right| + \left| (kx(s) - y(s))^{-(1+\alpha)} \right| \right) ds\\
&= O(kx(t_0) \pm y(t_0))^{-\alpha}
\end{aligned}
$$

i.e. they can be made arbitrarily small if $kx(t_0) \pm y(t_0)$ were taken large enough. This ensures that $\hat{p} \in U_\delta$ (with the same $\delta$), as required. To prove the contraction one may see from equations obtained by the differentiation of (5.7) that the norm of the derivative of $S$ (with respect to the functions $(p_x, p_y) \in U_\delta$) is small. Indeed, while the derivatives of $(x(s), y(s))$ with respect to $p$, denoted below by $X(s), Y(s)$, grow linearly in time, the $n$-th derivatives of $V$ decay as $t^{-n-\alpha}$ (here again we use that even under small deviations $(x(s), y(s))$ are bounded away from the boundary, so $kx(s) \pm y(s)$ grow with non-zero velocity as $s \to +\infty$).

Hence, the terms

$$\left|\int_t^{t_1} \frac{\partial^2}{\partial x^2} V_\varepsilon(x(s),y(s))X(s)ds\right|, \left|\int_t^{t_1} \frac{\partial^2}{\partial y^2} V_\varepsilon(x(s),y(s))Y(s)ds\right|,$$
$$\left|\int_t^{t_1} \frac{\partial^2}{\partial x\partial y} V_\varepsilon(x(s),y(s))X(s)ds\right|, \left|\int_t^{t_1} \frac{\partial^2}{\partial x\partial y} V_\varepsilon(x(s),y(s))Y(s)ds\right|$$

are small, which proves the contraction. The boundedness of the higher-order derivatives of $S$ is guaranteed by assumption (5.5) at the higher values of $n$. □

By the Banach principle of contraction mappings, the operator $S$ has a unique fixed point $(x(t),y(t),p(t))_{t\in[t_0,t_1]}$ and it depends on $x(t_0),y(t_0),p(t_1)$ smoothly. In fact, the above estimates show that the derivative of $S$ can be made as small in norm as we need, provided $kx(t_0)\pm y(t_0)$ are large enough (it decays as $O(kx(t_0)\pm y(t_0))^{-\alpha})$). Hence, the derivative $\frac{\partial p(t)}{\partial p(t_1)}$ is close to identity.

By construction, this fixed point is a solution of (5.6), i.e. it gives a trajectory of (5.2). The corresponding value of $p(t_0)$ is defined uniquely by $x(t_0)$, $y(t_0)$ and $p(t_1)$, moreover it depends on this data smoothly for any $t_0,t_1$, including $t_1=\infty$. Since $\left|\frac{\partial p(t_0)}{\partial p(t_1)}\right| \neq 0$, it follows that the exiting value of the vector of momenta $p(t_1)$ depends, in turn, smoothly on $x(t_0)$, $y(t_0)$ and $p(t_0)$.

Denote by $t_1(\varepsilon)$ the exit time of the trajectory from the region $\bar{C}_\varepsilon$, where the rescaled system is defined, so that $x(t_1)=(R-x_\varepsilon)/\delta$. Since $\dot{x}(t_1)=p_x(t_1)\neq 0$, it follows that $t_1$ is defined from this condition uniquely and depends continuously on $\varepsilon$ and smoothly on the initial conditions, where, at $\varepsilon=0$ we define $t_1=+\infty$. In any case we have finally that $p(t_1(\varepsilon))$ depends smoothly on initial conditions and continuously on $\varepsilon$. □

The existence of an asymptotic angle implies that the Hamiltonian trajectory moves finally in a wedge which is close to its asymptotic angle, but it does *not* necessarily approach a straight line. Hence, a more precise definition of the trajectory asymptotic is needed, as well as a precise definition of the asymptotic vertical shift $\eta_{out}$:

**Lemma 2.** *There exists a function $\mathcal{F}(x,\varphi)=x(\tan\varphi+o(1))$ such that the trajectory of $(x(0),y(0),p_x(0),p_y(0))\in\bar{C}_\varepsilon$ is of the following asymptotic form as $\varepsilon\to 0$ and $t\to\infty$:*

$$y(t)=\mathcal{F}(x(t),\varphi_{out})+\eta_{out}+\dots \tag{5.8}$$

*where $\varphi_{out}$ is the asymptotic outgoing angle of the trajectory and the dots stand for terms which go to zero in this limit. Similarly, as $\varepsilon\to 0$ and $t\to-\infty$,*

$$y(t)=\mathcal{F}(x(t),-\varphi_{in})+\eta_{in}+\dots. \tag{5.9}$$

*Furthermore, at $\varepsilon=0$, $(\varphi_{in},\eta_{in})$ defines the trajectory $(x(t),y(t))$ uniquely.*

*Proof.* By proposition 1 the asymptotic values of the velocity are well-defined. However, to obtain the asymptotic formulas for the behavior of the coordinates $(x,y)$ as $t\to\pm\infty$ we need more information about the derivatives of the solution of (5.7). Let us prove that for a fixed value of $\varphi_{out}$ the derivative $(X,Y,P_x,P_y)\equiv\frac{\partial(x,y,p_x,p_y)}{\partial(x(t_0),y(t_0))}$ is bounded for all times, moreover it has a finite limit as $t\to+\infty,\varepsilon\to+0$. Indeed, the solution of (5.7), as a fixed point of a contracting operator, can be found as the limit of successive approximations computed as follows: substitute the $m$-th approximation in the right-hand side of (5.7) and the result will be the $(m+1)$-th approximation. The approximations converge with all the derivatives with respect to $(x(t_0),y(t_0),p_x(t_1),p_y(t_1))$. We will show now that the boundedness and the convergence to a limit of $(X,Y,P_x,P_y)$ hold for all successive approximations uniformly, and hence this remains true for the trajectory defined by (5.7) (as it is the limit of the successive approximations).

By differentiation of (5.7) we get:

$$(X_{m+1}(t),Y_{m+1}(t)) = \begin{pmatrix} 1 & 0 \\ 0 & 1 \end{pmatrix} + \int_{t_0}^{t} (P_x(s),P_y(s))ds,$$
$$P_x(t) = \int_t^{t_1} \left( \frac{\partial^2 V_\varepsilon(x(s),y(s))}{\partial x^2} X_m(s) + \frac{\partial^2 V_\varepsilon(x(s),y(s))}{\partial x \partial y} Y_m(s) \right) ds$$
$$P_y(t) = \int_t^{t_1} \left( \frac{\partial^2 V_\varepsilon(x(s),y(s))}{\partial x \partial y} X_m(s) + \frac{\partial^2 V_\varepsilon(x(s),y(s))}{\partial y^2} Y_m(s) \right) ds$$

Using the decay rate of the potential and its derivatives (5.5), and the fact that the distance from the trajectory to the boundaries grows with a non-zero velocity, we immediately obtain that if $(X_m,Y_m)$ are bounded, then $(P_x(t),P_y(t)) = O(t^{-\alpha})$. Hence, the integral term in the first equation here is small (at large $t_0$) and convergent as $t \to +\infty$, which proves the claim.

Note that we have also shown that when we start with sufficiently large values of $x(t_0)$, the matrix $(X,Y) \equiv \begin{pmatrix} X_1 & X_2 \\ Y_1 & Y_2 \end{pmatrix}$ is close to identity and $P_x$, $P_y$ are close to zero. It is also true ( and it can be checked in the same manner) that all the derivatives of the solution of (5.7) which include at least one differentiation with respect to $(x(t_0),y(t_0))$ have finite limits as $t \to +\infty$.

Now, for each $\overline{\varphi}_{out}$, fix a trajectory $\overline{q}(t) = (\overline{x}(t),\overline{y}(t),\overline{p}_x(t),\overline{p}_y(t))$ with an asymptotic exit angle $\overline{\varphi}_{out}$. Define the function $\mathcal{F}(x,\overline{\varphi}_{out})$ as $\overline{y}(t) = \mathcal{F}(\overline{x}(t),\overline{\varphi}_{out})$. Now, consider another trajectory $q(t)$ which have the same asymptotic exit angle $\overline{\varphi}_{out}$. Let $x(t_0) = \overline{x}(t_0)$, where $p_x(t_0) > 0$, $x(t_0)$ is sufficiently large, and $y(t_0) = \overline{y}(t_0) + \Delta y_0$ where $\Delta y_0 \neq 0$. Then, $y(t) = \overline{y}(t) + \int_0^{\Delta y_0} Y_2(t) dy_0$ and $x(t) = \overline{x}(t) + \int_0^{\Delta y_0} X_2(t) dy_0$. It follows that

$$y(t) - \mathcal{F}(x(t),\overline{\varphi}_{out}) = \mathcal{F}(\overline{x}(t),\overline{\varphi}_{out}) + \int_0^{\Delta y_0} Y_2 dy_0 - \mathcal{F}(\overline{x}(t) + \int_0^{\Delta y_0} X_2 dy_0, \overline{\varphi}_{out}).$$

Since $X_2,Y_2$ are bounded and have a finite limit as $t \to +\infty$, and, by construction $\frac{\partial \mathcal{F}}{\partial x} = p_y/p_x$, it follows that $y(t) - \mathcal{F}(x(t),\overline{\varphi}_{out})$ has indeed a finite limit, defined to be $\eta_{out}$ as stated in the lemma.

In other words, all the orbits with a given value of $\varphi_{out}$ (recall that here $|\varphi_{out}| < \frac{\theta}{2}$) have the same asymptotic behavior as $t \to +\infty$ up to bounded terms. Namely, as $t \to +\infty$,

$$y(t) = \mathcal{F}(x(t), \varphi_{out}) + \eta_{out} + \dots \tag{5.10}$$

where the dots stand for the terms which tend to zero as $t \to +\infty$. The constant parameter $\eta_{out}$ distinguishes between different trajectories with the same values of $\varphi_{out}$.

If we fix $\varphi_{out}$ and some sufficiently large initial value $x(t_0)$, then by differentiating (5.10) with respect to the initial value $y(t_0)$ we obtain

$$\frac{\partial \eta_{out}}{\partial y(t_0)} = \frac{\partial y(t)}{\partial y(t_0)} - \frac{\partial \mathcal{F}}{\partial x} \frac{\partial x(t)}{\partial y(t_0)} + \dots.$$

As we mentioned, the quantities $Y_2(t,t_0) \equiv \frac{\partial y(t)}{\partial y(t_0)}$, $X_2(t,t_0) \equiv \frac{\partial x(t)}{\partial y(t_0)}$ have a finite limit as $t \to +\infty$, and $\frac{\partial \mathcal{F}}{\partial x}(x(t),\varphi_{out}) \equiv \overline{p}_y(t)/\overline{p}_x(t)$ has a finite limit as well. Moreover, $Y_2$ is close to 1 and $X_2$ is close to zero at all $t \geq t_0$, provided $t_0$ is large. Hence, the derivative $\frac{\partial \eta_{out}}{\partial y(t_0)}$ is close to 1 as well, i.e. it is bounded away from zero. It follows that given $\varphi_{out}$, the value of $\eta_{out}$ defines the orbit uniquely at $\varepsilon = 0$. The case $t \to -\infty$ is treated absolutely analogously. In fact, formula (5.9) follows from (5.8) by reversibility of the system: the problem is symmetric with respect to the transformation $t \leftrightarrow -t$, $\varphi_{out} \leftrightarrow -\varphi_{in}$. □

By the closeness of the trajectories to the billiard trajectories $\lim_{x\to+\infty} \mathcal{F}(x,\varphi_{out})/x = \tan\varphi_{out}$. One can show that $\mathcal{F}$ is linear with respect to $x$, provided we take $\alpha > 1$ in (5.5), but we do not need to assume this.

5.3. **The scattering function.** It follows from proposition 1 and lemma 2 that for sufficiently small $\varepsilon$ the trajectories of the system define a map

$$(\varphi_{in},\eta_{in}) \mapsto (\varphi_{out},\eta_{out}).$$

At $\varepsilon = 0$ the values of $\varphi_{in,out}$ and $\eta_{in,out}$ are taken at $t = \pm\infty$, i.e. they define the asymptotic behavior of the orbit. At small $\varepsilon > 0$ we define $\varphi_{out}$ as $\arctan \frac{p_y(t_1)}{p_x(t_1)}$ and $\eta_{out}$ as $y(t_1) - \mathcal{F}(x(t_1),\varphi_{out})$ where $t_1$ is the moment when the orbit exits the region $\bar{C}_\varepsilon$ (i.e. $x(t_1) = (R - x_\varepsilon)/\delta$ – recall that we are working here in the rescaled variables). The values of $\varphi_{in}$ and $\eta_{in}$ are defined analogously as $-\arctan \frac{p_y(t_{-1})}{p_x(t_{-1})}$ and $y(t_{-1}) - \mathcal{F}(x(t_{-1}),-\varphi_{in})$ at the moment $(t_{-1})$ the orbit enters $\bar{C}_\varepsilon$. We will be particularly interested in the dependence of $\varphi_{out}$ on $\eta_{in}$ at a given $\varphi_{in} \neq \pm\frac{\theta}{2}$. Denote

$$\varphi_{out} = \Phi^{\varepsilon}(\varphi_{in},\eta_{in}), \eta_{out} = \Psi^{\varepsilon}(\varphi_{in},\eta_{in}). \tag{5.11}$$

Summarizing, we have proved so far:

**Lemma 3.** *The functions $\Phi$, $\Psi$ are continuous in their arguments and $\varepsilon$, and they are $C^r$-smooth with respect to $(\varphi,\eta)$ when $\varphi_{in},\varphi_{out} \neq \pm\frac{\theta}{2}$.*

We will call $\Phi^{\varepsilon=0}$ the *scattering function*. It seems hardly possible to find an explicit expression for the scattering function in terms of the potential $V_0$. However, we can obtain some qualitative information about it. In particular, we prove the following result which shows that the billiard scattering functions $\Phi_-(\varphi_{in};\theta), \Phi_+(\varphi_{in};\theta)$ supply asymptotic information regarding $\Phi^0$.

**Lemma 4.** *For any $\varphi_{in} \in (-\frac{\theta}{2},\frac{\theta}{2})$ :*

$$\lim_{\eta_{in}\to-\infty} \Phi^0(\varphi_{in},\eta_{in}) = \Phi_-(\varphi_{in};\theta) \tag{5.12}$$

$$\lim_{\eta_{in}\to\infty} \Phi^0(\varphi_{in},\eta_{in}) = \Phi_+(\varphi_{in};\theta). \tag{5.13}$$

*Proof.* Fixing $\varepsilon$, and hence a cross-section $x = R_\varepsilon$, we may take $|\eta^{\varepsilon}_{in}|$ sufficiently large and guarantee that $\left|\varphi_{in} + \arctan \frac{y^{\varepsilon}_{in}}{R_\varepsilon}\right| >$ const $> 0$ (independent of $\varepsilon$). Then, according to lemma 8 (see appendix), the value of $\varphi^{\varepsilon}_{out}$ is indeed close to one of the billiards exit directions, i.e. to $\Phi_+(\varphi_{in})$ if the upper boundary is approached first (this corresponds to $\eta_{in} \sim +\infty$) or to $\Phi_-$ otherwise. Taking the limit $\varepsilon \to 0$, corresponding to $t \to -\infty$, and using lemma 3 proves the result. □

The continuity of the scattering function and the above result regarding its limiting values imply:

**Corollary 1.** *For any $\varphi_{in} \in (-\frac{\theta}{2},\frac{\theta}{2})$ the range $\mathcal{R}(\varphi_{in})$ of the scattering function $\Phi^0(\varphi_{in},\eta_{in})$ includes the interval $[\Phi_{-(-1)^{N_\theta}}(\varphi_{in};\theta), \Phi_{(-1)^{N_\theta}}(\varphi_{in};\theta)]$ :*

$$[\Phi_{-(-1)^{N_\theta}}(\varphi_{in};\theta), \Phi_{(-1)^{N_\theta}}(\varphi_{in};\theta)] \subseteq \mathcal{R}(\varphi_{in}) \subseteq [-\frac{\theta}{2},\frac{\theta}{2}]. \tag{5.14}$$

5.3.1. *The range of the scattering function-an example.* It is important to note that the left inclusion in 5.14 can be strict, i.e. the range $\mathcal{R}(\varphi_{in})$ can be larger than the interval between the limit values $\Phi_{\pm}$, because the function $\Phi^0(\varphi_{in},\eta_{in})$ need not be monotone (at least for some potentials). Indeed, consider for example a potential which is *symmetric with respect to reflection along the x-axis*, e.g.:

$$V_0(x,y) = \frac{1}{(kx-y)^\alpha} + \frac{1}{(kx+y)^\alpha}.$$

Take $\theta = \pi/n$ and $\varphi_{in} = 0$. Then, $\Phi_+ = \Phi_- = 0$ (see (3.1)). Hence, to show that the range of the function $\Phi^0$ at $\varphi_{in} = 0$ is not $\{0\}$ it is enough to show that it is not a constant, for example that $\frac{\partial \Phi^0}{\partial \eta}\Big|_{(0,\eta)} \neq 0$ at some $\eta$. Take $\eta = 0$, which corresponds to considering the trajectory which enters the corner along the bi-sector. Then, since $\dot{p}_y = 0$, the corresponding orbit of (5.2) is given by the equation:

$$y = 0,\ \frac{1}{2}\dot{x}^2 + V_0(x,0) = \frac{1}{2}\dot{x}^2 + \frac{2}{(kx)^\alpha} = 1.$$

If $\Phi^0(0,\eta)$ were a constant, then $\frac{\partial}{\partial \eta}\Phi^0(0,0) = 0$, namely solutions with nearby initial conditions with zero vertical momentum would end up with zero vertical momentum. We check that this is impossible for some values of $\alpha$ and $k$. Consider the equations for $Y(t) = \frac{\partial y}{\partial \eta}$. Since $\frac{\partial^2 V_0(x,y)}{\partial x \partial y}|_{y=0} = 0$ we get:

$$\ddot{Y} + \frac{\partial^2 V_0(x,y)}{\partial y^2}|_{y=0} Y = 0,$$

i.e. the condition $\frac{\partial}{\partial \eta}\Phi^0(0,0) = 0$ is equivalent to the existence of a non-trivial solution (as $Y(0) \approx 1 \neq 0$) to the following linear problem:

$$\ddot{Y} + \frac{\alpha(\alpha+1)}{(kx(t))^{\alpha+2}} Y = 0, \quad \dot{Y}(+\infty) = \dot{Y}(-\infty) = 0 \tag{5.15}$$

where $x(t) = x(-t)$ solves, for $t \geq 0$:

$$\frac{dx}{dt} = \sqrt{2 - \frac{4}{(kx)^\alpha}}, \qquad x(0) = \frac{2^{\frac{1}{\alpha}}}{k}.$$

It is easy to see that every such solution must be bounded and either even or odd. One may, however, check (we did it only numerically) that for $k = \tan\frac{\pi}{6}$ and $\alpha = 2$ (for which $x(t) = \frac{\sqrt{(2+2t^2k^2)}}{k}$) both the even and odd fundamental solutions to the $Y$ equations are unbounded. This demonstrates that $\Phi^0(\varphi_{in},\eta)$ is non-constant at $\varphi_{in} = 0, \theta = \frac{\pi}{3}$ near $\eta = 0$, hence $\mathcal{R}(\varphi_{in}) \neq \{0\} = [\Phi_+(0;\frac{\pi}{3}), \Phi_-(0;\frac{\pi}{3})]$.

More generally we conjecture:

**Conjecture 1.** *The spectrum of the values of* $\alpha$ *for which (5.15) has a localized solution is discrete.*

Provided this conjecture is true, for almost every $\alpha$ the function $\Phi^0(\varphi_{in},\eta)$ has extrema at $\varphi_{in} = 0$, $\theta = \frac{\pi}{n}$) and hence, for every close $\varphi_{in}$ and $\theta$. It is unclear yet how general this property is.

5.4. **The corner map.** Let us now proceed to the study of the behavior of the original system (2.1) near a corner. So, we return to the non-scaled coordinates $(x,y)$. Take a cross-section $x = R$ for some small $R > 0$. By the Proposition 1, every orbit which enters the region $x \leq R$ will eventually leave it crossing the cross-section again, hence, the corner return map:

$$\begin{aligned} T^{\varepsilon}_{cor} &: \Sigma^- \to \Sigma^+ \\ &: (y_{in}, \varphi_{in}) \mapsto (\ y^{\varepsilon}_{out}, \varphi^{\varepsilon}_{out}) \end{aligned}$$

is well defined. Here $y$ is the coordinate of the point of intersection with the cross-section and $\varphi$ defines the direction of the velocity at the cross-section as in figure 1. This is exactly the corner map that was defined in Theorem 1.

Let us make $R$ a function of $\varepsilon$ which tends to zero so slow that all the previous results, which we obtain for fixed $R$, are still valid. We will also assume that the scaling constants $\delta, x_\varepsilon, y_\varepsilon$ from (5.1) tend to zero faster than $R(\varepsilon)$. The following lemma is the main result of this section:

**Lemma 5.** *When* $|\varphi_{in}| < \frac{\theta}{2}$, *the corner map can be written as*

$$y^{\varepsilon}_{out} = R(\tan\varphi^{\varepsilon}_{out} + y^*_{\varepsilon}(\varphi_{out})) + \delta\Psi^{\varepsilon}(\varphi_{in}, \eta_{in}), \qquad \varphi^{\varepsilon}_{out} = \Phi^{\varepsilon}(\varphi_{in}, \eta_{in}) \tag{5.16}$$

*where*

$$\eta_{in} = \frac{y^{\varepsilon}_{in} + R(\tan\varphi_{in} - y^*_{\varepsilon}(-\varphi_{in}))}{\delta} \tag{5.17}$$

*The coefficient* $y^*_{\varepsilon}(\varphi)$ *is a smooth function of* $\varphi$ *which tends to zero as* $\varepsilon \to 0$ *along with all its derivatives.*

*Proof.* This follows from the construction of the scattering function (see (5.11)) and the asymptotic behavior of the solutions in the rescaled coordinates (see lemma 2 and formula (5.1)):

$$\begin{aligned} y^{\varepsilon}_{out} &= y_\varepsilon + \delta\mathcal{F}\left(\frac{R - x_\varepsilon}{\delta}, \varphi_{out}\right) + \delta\eta_{out} \\ &= y_\varepsilon + (R - x_\varepsilon)\tan\varphi_{out} + \delta\eta_{out} + o(R - x_\varepsilon) \\ &= R\tan\varphi_{out} + \delta\eta_{out} + o(R). \end{aligned}$$

By denoting the $o(R)$-term here as $Ry^*_{\varepsilon}$ we obtain

$$y^{\varepsilon}_{out} = R(\tan\varphi_{out} + y^*_{\varepsilon}(\varphi_{out})) + \delta\eta_{out}. \tag{5.18}$$

Now formula (5.16) follows immediately from (5.11). Relation (5.17) follows from (5.18) by the reversibility of the system (recall that the problem is invariant with respect to the transformation $t \leftrightarrow -t$, $\varphi_{in} \leftrightarrow \varphi_{out}$. □

Recall that in Theorem 1 we have shown that if a polygon within a billiard is a limit of some trajectory of (2.1), and if it enters a corner and leaves it with the angles $\phi_{in}$ and $\phi_{out}$, then $\phi_{out} \in I(\phi_{in})$. It follows from the above lemma that $I(\phi_{in}) = \mathcal{R}(\phi_{in})$, the range of the scattering function $\Phi^0$.

## 6. Hamiltonian flows near corner polygons.

After understanding the properties of the corner map $T^{\varepsilon}_{cor}$ (from $\Sigma^-$ to $\Sigma^+$) we are in a position to combine it with the external return map $T^{\varepsilon}_{ext}$ (from $\Sigma^+$ to $\Sigma^-$) and establish when corner polygons correspond to a limit of periodic orbits of the Hamiltonian flow. It turns out that one requirement is the following non-degeneracy condition:

**Definition 4.** *A corner polygon of the billiard is said to be* ***non-degenerate*** *if $\phi_{out} \in \mathcal{R}(\phi_{in})$ and infinitesimally small changes in $\phi_{out}$ change the return position of the trajectory so that the corner is missed.*

The external return map $T^{\varepsilon}_{ext}$ is defined by the trajectories on the cross-section $\{(x,y,\varphi)|x = R > 0\}$ near the corner, and it maps a small neighborhood of $(\varphi, y) = (\phi_{out}, y_{out} = R\tan\phi_{out})$ to a small neighborhood of $(\phi_{in}, y_{in} = -R\tan\phi_{in})$ (see Theorem 1 for more details). As above, we will take $R$ tending sufficiently slowly to zero as $\varepsilon \to +0$. Since the corner polygon has a finite number of regular reflections at $x > R$, the corresponding external return map by the billiard flow, $T^{0}_{ext}$, is smooth, and the Hamiltonian return map, $T^{\varepsilon}_{ext}$, limits to it in the $C^r$ topology [19]. In particular, defining $B^{\varepsilon}$ to be the derivative matrix of the external return map, $B^{\varepsilon} = (T^{\varepsilon}_{ext})'(\phi_{out}, y_{out})$, we conclude that $B^{\varepsilon}$ has a finite limit $B^0$ as $\varepsilon \to +0$. By definition, a corner polygon is non-degenerate if and only if $B^0_{21} \neq 0$.

**Theorem 2.** *Consider a family $V(x,y;\varepsilon)$ of billiard-like potentials limiting to a billiard in $D$ and satisfying the scattering assumption and the corner scaling assumption. Assume $D$ has a non-degenerate corner polygon with ingoing and outgoing angles $(\phi_{in}, \phi_{out})$. Then, for sufficiently small $\varepsilon$, for every $\eta$ such that $\phi_{out} = \Phi^0(\phi_{in}, \eta)$ and $\frac{\partial}{\partial\eta}\Phi^0(\phi_{in}, \eta) \neq 0$, the Hamiltonian family has a hyperbolic periodic orbit which, as $\varepsilon \to 0$, limits to the billiard corner polygon.*

*Proof.* Let us consider the combined map of the external and corner return maps to $x = R$ in the vicinity of this orbit:

$$T^{\varepsilon}_{ext} \circ T^{\varepsilon}_{cor}:$$

$$\begin{pmatrix} \bar{\varphi} \\ \bar{y} \end{pmatrix} = T^{\varepsilon}_{ext}\begin{pmatrix} \Phi^{\varepsilon}(\varphi,\eta) \\ R(\tan\Phi^{\varepsilon}(\varphi,\eta) + y^*_{\varepsilon}(\Phi^{\varepsilon}(\varphi,\eta))) + \delta\Psi^{\varepsilon}(\varphi,\eta) \end{pmatrix}$$
$$= B^{\varepsilon}\begin{pmatrix} \Phi^{\varepsilon}(\varphi,\eta) - \phi_{out} \\ R(\tan\Phi^{\varepsilon}(\varphi,\eta) + y^*_{\varepsilon}(\Phi^{\varepsilon}(\varphi,\eta))) + \delta\Psi^{\varepsilon}(\varphi,\eta) - R\tan\phi_{out} \end{pmatrix}$$
$$+\begin{pmatrix} \phi_{in} + \mu(\varepsilon) \\ R\tan\phi_{in} + \nu(\varepsilon) \end{pmatrix} + \dots \tag{6.1}$$

where $\mu(\varepsilon), \nu(\varepsilon)$ are the Hamiltonian corrections to the billiard external return map (hence, by [19], their limit is 0 as $\varepsilon \to 0$). The dots stand for quadratic and higher order corrections to the linearized external return map, and

$$\eta = \frac{y + R(\tan\varphi - y^*_{\varepsilon}(-\varphi))}{\delta}.$$

Plugging this expression in the fixed point equation for (6.1) and taking the limit $\varepsilon \to 0$ (with $R \to 0$ slowly with $\varepsilon$) we obtain:

$$\begin{pmatrix} \varphi \\ 0 \end{pmatrix} = \begin{pmatrix} B_{11}(\Phi^0(\varphi,\eta) - \phi_{out}) + \phi_{in} \\ B_{21}(\Phi^0(\varphi,\eta) - \phi_{out}) \end{pmatrix} + \dots, \tag{6.2}$$

where the dots stand for terms quadratic (or of higher order) in $(\Phi^0(\varphi,\eta) - \phi_{out})$.

This system has a solution $(\eta^*,\ \phi_{in})$ where $\eta^*$ solves $\phi_{out} = \Phi^0(\eta^*;\phi_{in})$ (notice that the terms denoted by dots in (6.2) vanish at this point). Furthermore, the Jacobian of the system is given by:

$$B_{21}\frac{\partial}{\partial\eta}\Phi^0(\phi_{in},\eta^*)$$

which, by our assumptions, is nonzero. Hence, by the implicit function theorem the fixed point equations have a nearby solution in $(\varphi,\eta)$ which implies that the Hamiltonian flow has the corresponding periodic orbit.

To find the fixed point stability, we calculate the trace of the linearized mapping. In the limit of small $\varepsilon$, the trace is given by $\frac{1}{\delta}B_{21}\frac{\partial}{\partial\eta}\Phi^0(\eta^*;\phi_{in}) + o(\frac{1}{\delta})$. As $\varepsilon,\delta \to 0$, the absolute value of the trace is certainly larger than 2 (the Jacobian of the return map at a periodic orbit equals to 1 by symplecticity), which shows that the periodic orbit we have found is hyperbolic. □

In Theorem 1 we proved that if the corner polygon is acceptable ($\phi_{out} \in \mathcal{R}(\phi_{in})$), then there exists a special perturbation of the given billiard-like potential family which attains a periodic orbit which limits to the corner polygon as $\varepsilon \to 0$. Theorem 2 demonstrates that analyzing the behavior near the corners pays — if the corner polygon is non-degenerate and the scattering function at the corresponding $\eta$ value has no extremum, then the results of theorem 1 are correct without the need of any perturbation (and the periodic orbit is hyperbolic).

Now we want to analyze the birth of elliptic periodic orbits from the corner polygons. By Theorem 2 this could happen only when a specific relation between $\phi_{in}$ and $\phi_{out}$ exists: given $\phi_{in}$, the value of $\phi_{out}$ has to be a local extremum of the scattering function. Existence of such a corner polygon is a codimension-1 phenomenon, so if we want to obtain a robust picture, it is necessary to consider here at least a one-parameter family of billiard tables.

This means that we must introduce an additional parameter, $\gamma$, in the potential $V$. At $\varepsilon = 0$ the potential is singular, so we need to define exactly to which class our one-parameter perturbations belong.

**Definition 5.** *A family of billiard-like potentials $V(x,y;\varepsilon,\gamma)$ is called a* ***tame*** *perturbation of the billiard-like potential $V(x,y;\varepsilon,0)$ if the barrier functions $W$ do not depend on $\gamma$, the pattern functions $Q$, defined in some neighborhood of the open boundary arcs without the corners, are $C^r$-smooth with respect to $\gamma$ and the rescaled potentials $V_\varepsilon$ depend $C^r$-smoothly on $\gamma$ as well.*

**Definition 6.** *The tame family of billiard-like potentials $V(x,y;\varepsilon,\gamma)$ is called non-degenerate at a corner polygon if:*

(6.3)
$$\mu'(\gamma)(B^0_{11}\frac{\partial}{\partial\varphi}\Phi^0(\phi_{in},\eta^*,\gamma)-1)-\nu'(\gamma)B^0_{21}\frac{\partial}{\partial\varphi}\Phi^0(\phi_{in},\eta^*,\gamma)-B^0_{21}\frac{\partial}{\partial\gamma}\Phi^0(\phi_{in},\eta^*,\gamma)\Big|_{\gamma=0} \neq 0$$

*where $\nu(\gamma)$ and $\mu(\gamma)$ represent the shifts in the angle and $y$ coordinates of the external return map $T^0_{ext}(\gamma)$ (see ( 4.2) at $\varepsilon = 0$), $\Phi^0$ is the scattering function of $V_0(x,y;\gamma)$, and $\eta^*$ is such that $\phi_{out} = \Phi^0(\phi_{in},\eta^*,0)$.*

**Theorem 3.** *Consider a family of billiard-like potentials $V(x,y;\varepsilon)$ limiting to a billiard in a domain $D$ and satisfying the scattering assumption and the corner scaling assumption with a scaling parameter $\delta(\varepsilon)$. Assume $D$ attains an acceptable non-degenerate corner polygon with ingoing and outgoing angles $(\phi_{in},\phi_{out})$. Let $V(x,y;\varepsilon,\gamma)$ be a one-parameter tame perturbation of $V(x,y;\varepsilon)$, satisfying the non-degeneracy assumption. Then, for every $\eta^*$ such* *that $\phi_{out} = \Phi^0(\phi_{in}, \eta^*)$ is a strict extremum (i.e. $\frac{\partial}{\partial\eta}\Phi^0(\phi_{in}, \eta^*) = 0$, and $\frac{\partial^2}{\partial\eta^2}\Phi^0(\phi_{in}, \eta^*) \neq 0$) there exists a wedge of width $\delta^2(\varepsilon)$ in the $(\varepsilon, \gamma)$ parameter plane in which the Hamiltonian flow defined by the potential $V(x, y; \varepsilon, \gamma)$ has elliptic islands of size $O(\delta^2)$, where the islands limit to the billiard corner polygon as $\varepsilon \to 0$.*

*Proof.* Construct the return map as in Theorem 2 (see (6.1)), with the shifts $\nu$ and $\mu$ depending on $\gamma$ now; by construction $\nu(\varepsilon, \gamma) = 0, \mu(\varepsilon, \gamma) = 0$ at $(\varepsilon, \gamma) = (0, 0)$. Taking the limit as $\varepsilon \to 0$, the fixed point equation of the previous theorem becomes:

$$\begin{pmatrix} \varphi \\ 0 \end{pmatrix} = \begin{pmatrix} B_{11}(\Phi^0(\varphi, \eta, \gamma) - \phi_{out}) + \phi_{in} + \nu(0, \gamma) \\ B_{21}(\Phi^0(\varphi, \eta, \gamma) - \phi_{out}) + \mu(0, \gamma) \end{pmatrix} + \dots .$$

At $\eta = \eta^*$ this system has a solution $\mu = \nu = 0$, $\varphi = \phi_{in}$. Therefore, this system has a solution $(\varphi(\eta), \gamma(\eta))$ for every $\eta \approx \eta^*$ provided the Jacobian with respect to variations of $\varphi, \gamma$ does not vanish. This Jacobian is given by:

$$\mu'(\gamma)(B_{11}\frac{\partial}{\partial\varphi}\Phi^0(\phi_{in}, \eta^*, \gamma) - 1) - \nu'(\gamma)B_{21}\frac{\partial}{\partial\varphi}\Phi^0(\phi_{in}, \eta^*, \gamma) - B_{21}^0\frac{\partial}{\partial\gamma}\Phi^0(\phi_{in}, \eta^*, \gamma)$$

which by our assumption is non-zero at $\gamma = 0$. Hence, for every $\eta$ close to $\eta^*$ there exists $\gamma$ for which the map has a fixed point with the given value of $\eta$. The trace of the linearized map at this point is given by $\frac{1}{\delta(\varepsilon)}B_{21}\frac{\partial}{\partial\eta}\Phi^0(\phi_{in}, \eta, \gamma) + o(\frac{1}{\delta})$, which by our assumptions changes sign across $\eta \approx \eta^*$ (recall that $B_{21} \neq 0$ because the corner polygon is non-degenerate). Thus, for sufficiently small $\varepsilon$, there exists an interval of $\eta$ (hence, $\gamma$) values for which the trace varies in the interval $(-2, 2)$, and these values of trace of $\gamma$ correspond to elliptic (linearly stable) periodic orbit.

To prove the existence of islands the linear information is insufficient - we need to show that the coefficients of some of the nonlinear terms in the local return map do not vanish. We prove this by transforming the return map, by a series of symplectic transformations, to a map which is close to the conservative Hénon map. Then, we complete the proof by establishing that for small $\varepsilon$ a small change in the bifurcation parameter $\gamma$ causes the Hénon map bifurcation parameter to vary across a large interval which includes the interval for which the Hénon map has an island of stability.

Rewrite the explicit return map which may be computed as in (6.1) symbolically as:

$$\begin{pmatrix} \varphi' \\ \delta\eta' \end{pmatrix} = \begin{pmatrix} p(\varphi, \eta, \varepsilon, \gamma) \\ q(\varphi, \eta, \varepsilon, \gamma) \end{pmatrix} = \begin{pmatrix} \tilde{p}(\varphi, \eta, \varepsilon, \gamma) + \tilde{\nu}(\gamma) \\ \tilde{q}(\varphi, \eta, \varepsilon, \gamma) + \tilde{\mu}(\gamma) \end{pmatrix} \tag{6.4}$$

where $\tilde{\nu} = \nu + B_{11}(\Phi^0(\phi_{in}, \eta^*, \gamma) - \phi_{out})$, $\tilde{\mu} = \mu + B_{21}(\Phi^0(\phi_{in}, \eta^*, \gamma) - \phi_{out})$.

Let $\overline{\eta}$ solve the equation:

$$\frac{\partial}{\partial\eta}q(\phi_{in}, \overline{\eta}, \varepsilon, \gamma) = 0.$$

Since $\varepsilon = \gamma = 0$, $\eta = \eta^*$, $\varphi = \phi_{in}$ solves this equation, and since $\frac{\partial^2}{\partial\eta^2}q(\phi_{in}, \eta^*, 0, 0) = B_{21}\frac{\partial^2}{\partial\eta^2}\Phi^0(\eta^*, \phi_{in}, 0) \neq 0$, solution to this equation exists for all small $\varepsilon$ and $\gamma$. Now, consider the return map in the shifted coordinates:

$$\widetilde{\varphi} = \varphi - \phi_{in}, \quad \widetilde{\eta} = \eta - \overline{\eta}$$

which may be written in the following form:

$$\begin{pmatrix} \widetilde{\varphi}' \\ \delta\widetilde{\eta}' \end{pmatrix} = \begin{pmatrix} p(\phi_{in}, \overline{\eta}, \varepsilon, \gamma) - \phi_{in} + p_1(\widetilde{\varphi}, \widetilde{\eta}, \varepsilon, \gamma)\widetilde{\varphi} + p_2(\widetilde{\eta}, \varepsilon, \gamma)\widetilde{\eta} \\ q(\phi_{in}, \overline{\eta}, \varepsilon, \gamma) - \delta\overline{\eta} + q_1(\widetilde{\varphi}, \widetilde{\eta}, \varepsilon, \gamma)\widetilde{\varphi} + q_2(\widetilde{\eta}, \varepsilon, \gamma)\widetilde{\eta}^2 \end{pmatrix} \tag{6.5}$$

Symplecticity implies (recall that the equations were multiplied by $\delta$ in (6.4), and the symplectic density here is finite since $p_x$ is bounded away from zero):

$$\begin{vmatrix} p_1(\tilde\varphi,\ \tilde\eta,\varepsilon,\gamma)+p_{1\tilde\varphi}(\tilde\varphi,\ \tilde\eta,\varepsilon,\gamma)\tilde\varphi & p_2(\tilde\eta,\varepsilon,\gamma)+p_{1\tilde\eta}(\tilde\varphi,\ \tilde\eta,\varepsilon,\gamma)\tilde\varphi+p_{2\tilde\eta}(\tilde\eta,\varepsilon,\gamma)\tilde\eta \\ q_1(\tilde\varphi,\ \tilde\eta,\varepsilon,\gamma)+q_{1\tilde\varphi}(\tilde\varphi,\ \tilde\eta,\varepsilon,\gamma)\tilde\varphi & q_{1\tilde\eta}(\tilde\varphi,\ \tilde\eta,\varepsilon,\gamma)\tilde\varphi+q_{2\tilde\eta}(\tilde\eta,\varepsilon,\gamma)\tilde\eta^2+2\tilde\eta q_2(\tilde\eta,\varepsilon,\gamma) \end{vmatrix} = O(\delta).$$

Taking $\tilde\varphi=0$ we obtain:

$$p_2(\tilde\eta,\varepsilon,\gamma)q_1(0,\ \tilde\eta,\varepsilon,\gamma)=O(\delta,|\tilde\eta|) \tag{6.6}$$

Notice that

$$q_1\equiv q_1(0,0,0,0)=B_{21}\frac{\partial}{\partial\varphi}\Phi^0(\phi_{in},\eta^*,0)$$

By symplecticity of the corner map, its Jacobian is non-zero at any point. Hence, $\frac{\partial}{\partial\varphi}\Phi^0(\phi_{in},\eta^*,0)\neq 0$ (recall that $\frac{\partial}{\partial\eta}\Phi^0(\phi_{in},\eta^*,0)=0$). Thus, (6.6) implies $p_2(\tilde\eta,\varepsilon,\gamma)=O(\delta,|\tilde\eta|)$. Now, let us rescale these shifted coordinates:

$$\delta^2\hat\varphi=\tilde\varphi,\ \ \delta\hat\eta=\tilde\eta.$$

Plugging in (6.5) and dividing by $\delta^2$ gives:

$$\begin{pmatrix}\hat\varphi'\\ \hat\eta'\end{pmatrix}=\begin{pmatrix} m_1+p_1\hat\varphi+\hat p_2\hat\eta+\hat p_3\hat\eta^2+\dots\\ m_2+q_1\hat\varphi+q_2\hat\eta^2+\dots\end{pmatrix} \tag{6.7}$$

where

$$\begin{aligned} p_1&\equiv p_1(0,\ 0,0,0)=B_{11}\frac{\partial}{\partial\varphi}\Phi^0(\phi_{in},\eta^*,0)\\ q_2&\equiv q_2(0,\ 0,0)=B_{21}\frac{\partial^2}{\partial\eta^2}\Phi^0(\phi_{in},\eta^*,0)\\ m_1&=\frac{p(\phi_{in},\bar\eta,\varepsilon,\gamma)-\phi_{in}}{\delta^2}\\ m_2&=\frac{q(\phi_{in},\bar\eta,\varepsilon,\gamma)-\delta\bar\eta}{\delta^2}\end{aligned} \tag{6.8}$$

and the terms denoted by dots tend to zero as $\varepsilon\to 0$. As we rescaled the symplectic density, this map is symplectic, moreover:

$$\begin{vmatrix} p_1 & \hat p_2+2\hat p_3\hat\eta\\ q_1 & 2q_2\hat\eta\end{vmatrix}=1$$

hence

$$\hat p_2 q_1=-1,\quad 2q_2p_1=2\hat p_3 q_1.$$

With a slight abuse of notation, letting

$$\begin{aligned}\varphi&=q_1q_2\hat\varphi-p_1q_2\hat\eta-q_1q_2m_1+p_1q_2m_2-\frac{p_1}{2}\\ \eta&=-q_2\hat\eta-\frac{p_1}{2}\end{aligned}$$

and plugging these expressions in (6.7) we obtain a perturbation of the Hénon map (the dots here stand for the terms which tend to zero as $\varepsilon\to 0$):

$$\begin{pmatrix}\varphi'\\ \eta'\end{pmatrix}=\begin{pmatrix}\eta+\dots\\ a-\varphi-\eta^2+\dots\end{pmatrix} \tag{6.9}$$

with the bifurcation parameter:

$$a(\gamma,\varepsilon)=q_2(-q_1m_1+(p_1-1)m_2)-p_1+\frac{p_1^2}{4}. \tag{6.10}$$

From (6.8) and (6.4):

$$m_1(\varepsilon,\gamma) = \frac{\tilde{\nu}(\gamma)}{\delta^2} + \frac{\tilde{p}(\phi_{in},\overline{\eta},\varepsilon,\gamma) - \phi_{in}}{\delta^2}$$
$$m_2(\varepsilon,\gamma) = \frac{\tilde{\mu}(\gamma)}{\delta^2} + \frac{\tilde{q}(\phi_{in},\overline{\eta},\varepsilon,\gamma) - \delta\overline{\eta}}{\delta^2}.$$

It can be shown, using the expansion of $\overline{\eta}$ near $\eta^*$ that the second terms of the $m_i$s are of lower order in $\delta$ and that $\frac{\partial m_1}{\partial\gamma}|_{(0,0)} \approx \frac{\tilde{\nu}'(0)}{\delta^2}$, and $\frac{\partial m_2}{\partial\gamma}|_{(0,0)} \approx \frac{\tilde{\mu}'(0)}{\delta^2}$. Clearly $\tilde{\mu}(0) = \tilde{\nu}(0) = 0$. Hence, by taking $\delta^2 = o(\gamma)$, $a(\gamma,\varepsilon)$ can be made to run through an arbitrary large interval $[-A,B]$ as $\varepsilon \to 0$ provided $\frac{\partial}{\partial\gamma}\left(\delta^2 a\right) \neq 0$. Using (6.8), (6.10) and the assumption of the non-degeneracy of the corner polygon (i.e. $B_{21} \neq 0$) we obtain that this condition reduces to (6.3). Summarizing, we have shown that for sufficiently small $\varepsilon$ the return map is conjugate to a map which is close to the Hénon map, hence, it has elliptic islands on open interval of $\gamma$ values, as the Hénon map does. From the rescaling it is clear that the width of those intervals in $\gamma$ is $O(\delta^2)$ as is the width and height of these islands in the original phase space coordinates. □

It follows that if the billiard is dispersing and the billiard map has a Lyapunov exponent $\lambda$, then if the corner polygon has $n+1$ edges, the bifurcation coefficient $a$ in the resulting coefficient in the Hénon map is proportional to $\lambda^{2n}$ (since $p_1, q_{1,2} \propto \left|B_{ij}\right| = O(\lambda^n)$), and the transformation to the Hénon map includes scaling of $(\varphi, y)$ by factors proportional to $(\lambda^{2n}, \lambda^n)$ respectively. Hence the size of the islands, in both parameter space and phase space, decreases exponentially with the number of reflections, as expected.

## 7. Geometrically created elliptic orbits

We have seen (see section 3, table 1) that in many cases the billiard corner map takes a parallel ray and bends it non-monotonically. Hence, it appears natural to establish that in the smooth system this bending creates islands. One can foresee two logical possibilities here. The first one is that this bending creates extrema in the scattering function – the birth of elliptic islands in this case was analyzed in the previous section. The second possibility is that the scattering function is monotone. In this case the bending of the parallel beam (hence - elliptic orbits) should occur in the region where the behavior of the system near the corner matches the billiard limit, i.e. at large values of $\eta$. The values $\eta = \pm\infty$ correspond to $\phi_{out} = \Phi_\pm(\phi_{in})$, and this is the case which we consider in Theorem 4 below (we formulate it only for the case $\phi_{out} = \Phi_+(\phi_{in})$; the case $\phi_{out} = \Phi_-(\phi_{in})$ is treated in a symmetric way).

**Theorem 4.** *Consider a nondegenerate corner polygon with $\phi_{out} = \Phi_+(\phi_{in})$. Assume that the scattering function is monotone at large positive $\eta$, and denote $\sigma = \mathrm{sign}(\frac{\partial}{\partial\eta}\Phi^0(\phi_{in},\eta))$ at large $\eta$. If*

$$(7.1)\qquad (-1)^{n_+ +1}\left(B_{11} + 2B_{21}\cos\phi_{in}\sum\frac{\kappa_{(-1)^{j+1}}}{\cos\alpha_j} + B_{22}\frac{\cos\phi_{in}}{\cos\phi_{out}}\right)\mathrm{sign}(B_{21})\sigma < 2$$

*where $\alpha_j$ are given by (3.6), and $\kappa_\pm$ is the curvature on the upper/lower arcs of the corner, then, for sufficiently small $\varepsilon$ an elliptic periodic orbit is produced by this billiard corner polygon.*

*Proof.* Consider a tame embedding family of billiard potentials $V(\cdot;\varepsilon,\mu,\nu)$. Below, we prove that for any such family there exists an interval of $\eta$ values, $(\eta_-^\varepsilon, \eta_+^\varepsilon)$ with $\eta_\pm^\varepsilon \to \infty$ as $\varepsilon \to 0$, for which the trace of the linearized return map to $\Sigma^-$ is in $(-2,2)$. Now, by lemma 4, at all $\eta$ sufficiently large the value of $\varphi_{out}$ will be close to $\Phi_+(\phi_{in})$. Therefore, from the proof of theorem 1, it is seen that we may always find $\mu(\varepsilon),\nu(\varepsilon)$ so that the Hamiltonian flow with the billiard potential $V(\cdot;\varepsilon,\mu(\varepsilon),\nu(\varepsilon))$ will have a periodic orbit with $\eta_{in}\in(\eta_-^\varepsilon,\eta_+^\varepsilon)$, namely an elliptic periodic orbit is produced.

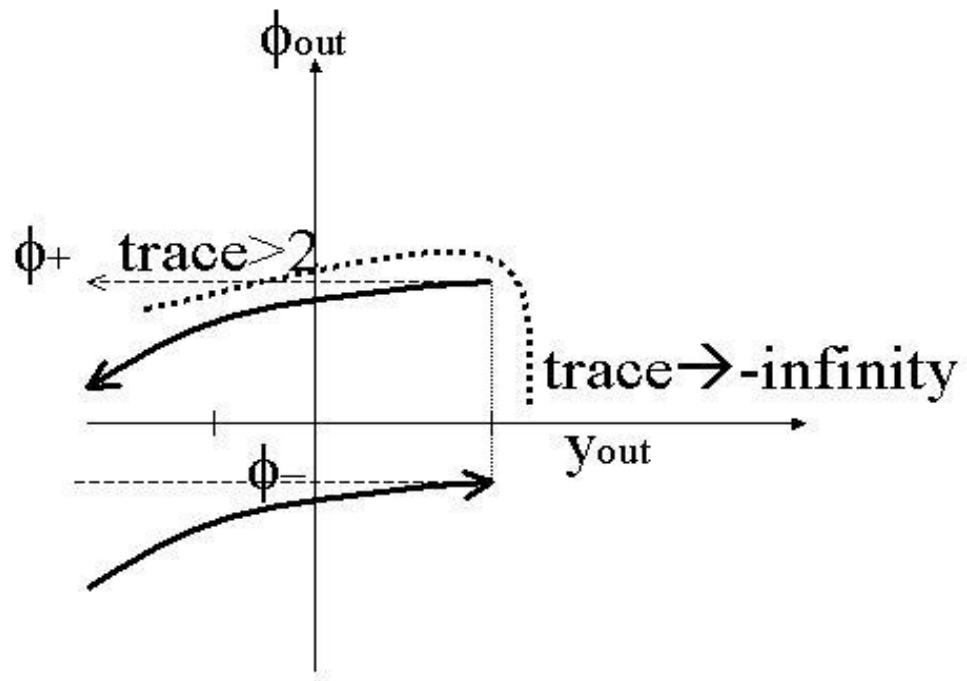

FIGURE 6. The Hamiltonian action on a parallel ray with non-monotonic behavior.

Now we prove that there is an interval $(\eta_-^\varepsilon,\eta_+^\varepsilon)$ with $\eta_\pm^\varepsilon\to\infty$ as $\varepsilon\to 0$, on which the trace is in $(-2,2)$. Fixing $\eta_{in}$ and letting $\varepsilon\to+0$, the trace of the derivative of the Poincaré map computed for this trajectory will be given, as in Theorem 2 by $\frac{1}{\delta}B_{12}\frac{\partial}{\partial\eta}\Phi(\phi_{in},\eta_{in})+o(\frac{1}{\delta})$, so it will be close to plus or minus infinity depending on the sign of $B_{12}\sigma$. On the other hand, if we allow $\eta_{in}$ to tend to infinity sufficiently fast, our periodic orbit will be close to the corresponding billiard orbit and the Poincaré map of the Hamiltonian flow will be close to the Poincaré map of the billiard flow along with its derivatives (here we use again the fact that $\varphi_{out}$ will be close to $\Phi_+(\phi_{in})$). Therefore, at such $\eta_{in}$ the trace of the derivative of the Poincaré map will be close to that we have for the billiard map. So, in the limit $\varepsilon\to+0$ the trace equals to (see (3.5)):

$$T=\text{trace}\left((-1)^{n_++1}\begin{pmatrix}B_{11}&B_{12}\\B_{21}&B_{22}\end{pmatrix}\begin{pmatrix}1&2\cos\phi_{in}\sum^{\kappa}\frac{(-1)^{j+1}}{\cos\alpha_j}\\0&\frac{\cos\phi_{in}}{\cos\phi_{out}}\end{pmatrix}\right)$$
$$=(-1)^{n_++1}\left(B_{11}+2B_{21}\cos\phi_{in}\sum^{\kappa}\frac{(-1)^{j+1}}{\cos\alpha_j}+B_{22}\frac{\cos\phi_{in}}{\cos\phi_{out}}\right).$$

Due to continuous dependence on the initial conditions, to ensure the existence of elliptic orbits, we need to show that the interval spanned by these two limiting trace values intersects the interval $(-2,2)$, and this amounts to the condition 7.1. The $\eta$ values for which this intersection occurs are $(\eta_-^\varepsilon,\eta_+^\varepsilon)$. To see that these values are arbitrarily large as $\varepsilon\to 0$, notice again that for any fixed $\eta$, the trace $\frac{1}{\delta}B_{12}\frac{\partial}{\partial\eta}\Phi(\phi_{in},\eta_{in})+o(\frac{1}{\delta})$ is arbitrarily large in magnitude. □

**Corollary 2.** *Consider a* ***dispersing*** *billiard-like family, with a nondegenerate corner polygon satisfying $\phi_{out}=\Phi_+(\phi_{in})$. If $\Phi^0(\phi_{in},\eta)$ is monotone, then, for sufficiently small $\varepsilon$ an* *elliptic periodic orbit is produced by the billiard corner polygon if* $(\frac{\pi}{\theta}-\left[\frac{\pi}{\theta}\right]-\frac{1}{2})\theta=\varphi_+^c<\phi_{in}<\frac{\pi}{2}$.

*Proof.* Notice that for dispersing billiards all the elements of the matrix $B$ have the same sign (see e.g. [19]; our choice of the orientation of $\phi_{in,out}$ is, of course, important here), and that the absolute value of the trace $(T)$ of the linearized motion about any periodic orbit in a dispersive billiard is larger than 2. Hence, the inequality 7.1 is satisfied iff $\sigma=\text{sign}(\frac{\partial}{\partial\eta}\Phi^0(\eta,\phi_{in}))=(-1)^{n_+}$. Furthermore, when the scattering function is monotone (in fact, it is sufficient to assume it is monotone for $\eta>\eta_0$, where $\eta_0$ is, for example, the largest solution of $\Phi^0(\eta_0,\phi_{in})=\frac{\Phi_+(\phi_{in})+\Phi_-(\phi_{in})}{2}$), the sign of its derivative coincides with the sign of $(\Phi_+(\phi_{in})-\Phi_-(\phi_{in}))$ i.e. it is defined by the billiard geometry alone. Now, it may be checked that $\text{sign}(\Phi_+(\phi_{in})-\Phi_-(\phi_{in}))=(-1)^{N_\theta}$. Since $n_+(\phi_{in})=N_\theta$ when $\varphi_+^c<\phi_{in}<\frac{\theta}{2}$, see 3.3, the corollary is proved. □

See the table of section 3 and figures 4 and 6 for the geometrical interpretation of the above condition - it basically shows that when a shoulder is created because the direction of the jump is opposite to the monotonicity implied by the billiard dispersiveness an elliptic orbit is created.

## 8. Summary and Conclusions

We have developed a framework for dealing with smooth approximations to billiards with corners in the two-dimensional setting. Given a billiard with a corner polygon, we proved that the smooth Hamiltonian flow can have a nearby periodic orbit if and only if the corner polygon angles at the corner are acceptable. The criteria for a corner polygon to be acceptable depends both on the geometry at the corner and on the smooth potential behavior at the corners (which determines the scattering function). We proved the existence of an asymptotic scattering function, explained how it can be calculated and predicted some of its properties, yet we were not able to calculate it explicitly (this seems to be impossible in the general case because of nonintegrability). We constructed a fixed point equation which defines the periodic orbit of the smooth system, and proved that the periodic orbit of the smooth system is hyperbolic provided the billiard polygon orbit is acceptable and non-degenerate and the scattering function is not extremal there. We then proved that if the scattering function is extremal, an elliptic periodic orbit arises, and, furthermore, that the return map near this periodic orbit is conjugate to a map close to the Hénon map and therefore has elliptic islands. We have found from the scaling that the island size is typically algebraic in the smoothing parameter and exponential in the number of reflections of the polygon orbit. Finally, we have proved that some corner polygons always produce elliptic orbits, independent of the details of the billiard potential.

We have analyzed the limiting behavior for a given, fixed corner (fixed $\theta$ value). Recall that the nature of the billiard flow at the corner is highly sensitive to the numerical properties of $\theta$, with bifurcation points at $\theta_N=\frac{\pi}{N}$ and $\theta_N^*=\frac{\pi}{N+\frac{1}{2}}$. The influence of these bifurcations on the limiting Hamiltonian flow is yet to be studied - it may produce nontrivial dynamics (e.g. the analysis of section 5.3.1), which is especially relevant for small angles.

Now, consider a one parameter family of dispersing billiards $D_\gamma$. One would like to characterize the appearance of islands for sufficiently small $\varepsilon$ as a function of $\gamma$. It is clear that for sufficiently small $\varepsilon$ the only mechanism for creating islands is the behavior of the smooth system near singular orbits of the billiard, namely near tangent orbits and near orbits which enter a corner. Generically, if no special symmetries are imposed, $D_0$ has many near-tangent periodic orbits, but no tangent ones. We conjecture that for generic families, a small deformation of $D_0$ to $D_\gamma$, can make a near-tangent periodic orbit of period $n$ to a tangent one for some $\gamma$ of order $\lambda^{-n}$, where $n \gg 1$. This implies that for sufficiently small $\varepsilon$, very small (size $\delta_{\tan}(\varepsilon)\lambda^{-n}$) islands will appear in the Hamiltonian approximation to $D_\gamma$. On the other hand, we expect $D_0$ to have many corner polygons, and in particular corner polygons with only one edge - a minimizing cord (a segment emanating from one of the corners which has a straight angle reflection from the boundary). Generically, these corner polygons will have the angles $\phi_{in}$ and $\phi_{out}$ in general position, i.e. $\phi_{out}$ will not be an extremum of the scattering function for the given $\phi_{in}$. So, according to our results above (theorem 2) only a saddle periodic orbit can be born from any such polygon at sufficiently small $\varepsilon$. However, due to the transitivity, we can expect sufficiently long corner orbits for which $\phi_{out}$ will be close to the extremum of the scattering function. Hence, some small islands can be obtained from these orbits after $\gamma$ is tuned appropriately.

Note that in applications where one needs to tailor a billiard table with a given properties the idea of small perturbation of the billiard boundary is, in fact, irrelevant, so one can consider large changes in $\gamma$ as well. Then, producing low period tangent orbits or minimizing cords with any given values of $(\phi_{in}, \phi_{out})$ is very easy. In this way one can produce elliptic islands of a visible size in dispersing billiard-like potentials.

## ACKNOWLEDGMENT

We thank N. Fridman, N. Davidson and U. Smilansky for useful discussions. Support of the Weierstrass institute and the Weizmann Institute for mutual visits is greatly appreciated.

## 9. APPENDIX

Here we prove Proposition 1 which is needed for establishing the properties of the dynamics in the scaled equations of motion.

9.1. **Proof of Proposition 1:** Notice that by the scattering assumption every trajectory must come to the region of sufficiently large values of the (rescaled) coordinate $x$ as $t \to \pm\infty$. Hence, we focus on the analysis of the behavior of the rescaled Hamiltonian flow (5.2) at large $x$. First, we prove that for the orbits in $\bar{C}_\varepsilon$, staying at a large distance from the boundaries of $\bar{C}_\varepsilon$, the momenta are essentially preserved:

**Lemma 6.** *Consider a billiard-like potential family satisfying the corner scaling assumption. For large $L$, for any orbit such that $kx(s) - |y(s)| > L$ for all $s \in [0, \tau]$ we have*

$$||p(\tau) - p(0)|| = O(L^{-\alpha/2}).$$

*Proof.* Assume, first, that $kp_x(0) - |p_y(0)|$ is bounded away from zero:

$$|kp_x(0) - |p_y(0)|| > A(k+2)L^{-\alpha/2}$$

where $A$ is some sufficiently large constant and $\alpha$ reflects the assumed decay rate of $\nabla V_\varepsilon$ (see (5.5) ). Let $||p(s) - p(0)|| \le AL^{-\alpha/2}$ at $s \in [0, t]$ for some $t$. It follows, in particular, that $|kp_x(s) - |p_y(s)|| \ge AL^{-\alpha/2}$ for all $s \in [0, t]$. There are two possibilities here. First, if $kp_x(s) - |p_y(s)| \ge AL^{-\alpha/2}$, then the distance to both boundaries grows at least linearly (with the velocity not less than $AL^{-\alpha/2}$), so we have the following estimate

$$||\nabla V_\varepsilon(x(s), y(s))|| \le 2K(L + AL^{-\alpha/2}s)^{-(1+\alpha)}.$$

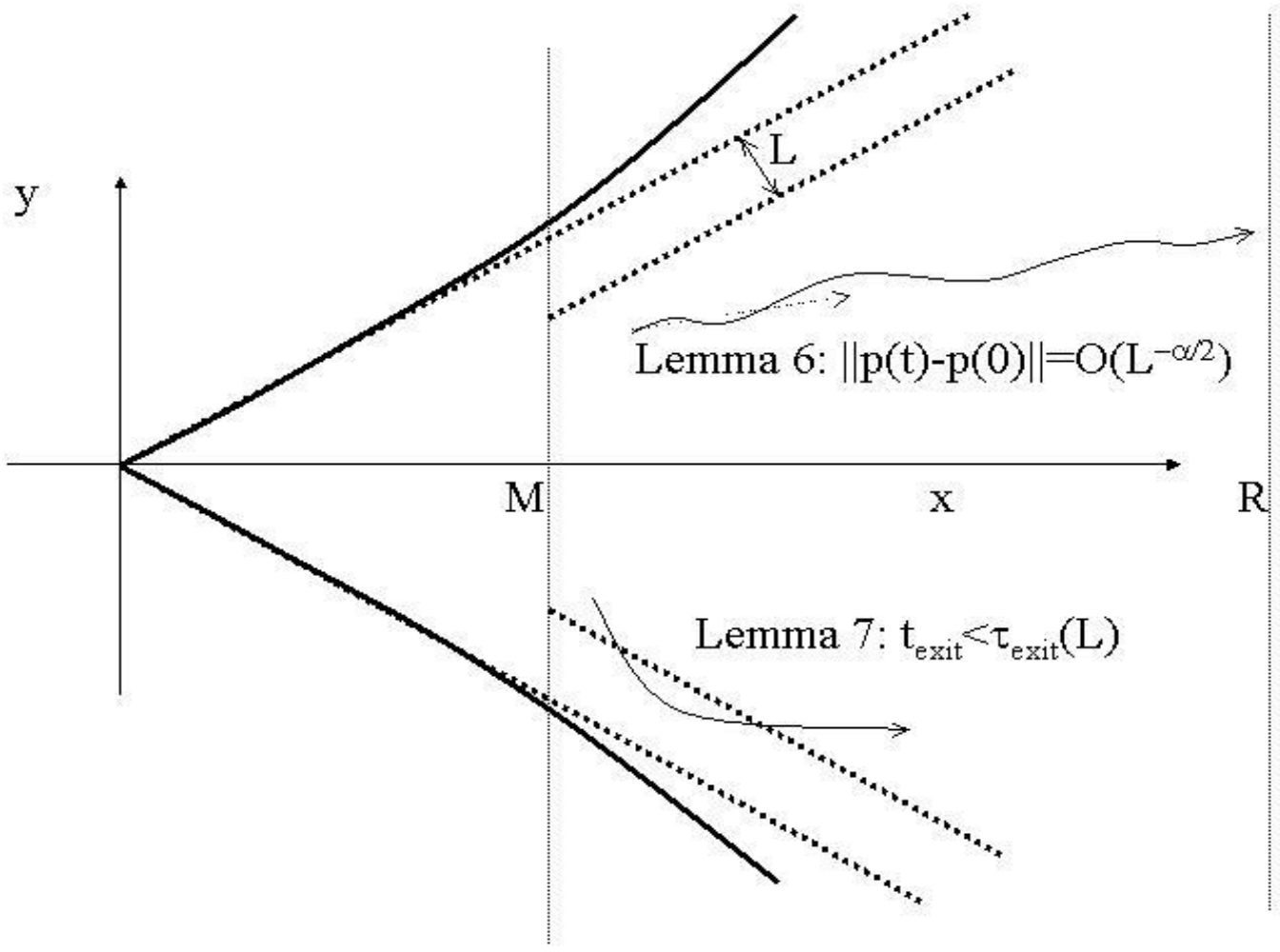

FIGURE 7. Geometry of the boundary layers.

By (5.2)

$$p(t) = p(0) - \int_0^t \nabla V_\varepsilon(x(s), y(s)) ds. \tag{9.1}$$

This gives us $||p(t) - p(0)|| \leq \frac{2K}{\alpha A} L^{-\alpha/2}$. Choosing $A > \sqrt{\frac{2K}{\alpha}}$, it follows that $||p(t) - p(0)|| < AL^{-\alpha/2}$ (with strict inequality). Thus, the inequality domain may be extended for all $t$, which proves the lemma.

The second possibility is $kp_x(s) - |p_y(s)| \leq -AL^{-\alpha/2}$. In this case, the distance to one of the boundaries increases and the distance to the other one decreases. Say, if $kp_x(s) - p_y(s) \leq -AL^{-\alpha/2}$, then the distance to the upper boundary decreases with velocity of at least $AL^{-\alpha/2}$ and the distance to the lower boundary increases (linearly in time, as well). We have the following estimate

$$||\nabla V_\varepsilon(x(s), y(s))|| \leq K((L + AL^{-\alpha/2}s)^{-(1+\alpha)} + (L + AL^{-\alpha/2}(t-s))^{-(1+\alpha)}$$

which, by (9.1), again gives us $||p(t) - p(0)|| < AL^{-\alpha/2}$ with the margin of safety.

In the remaining case, where $kp_x - |p_y|$ is, initially, $O(L^{-\alpha/2})$-close to zero, if it eventually would deviate from zero to the distance $A(k+2)L^{-\alpha/2}$, then the above arguments show that the further change in $p$ cannot exceed $AL^{-\alpha/2}$. Thus, $|kp_x - |p_y||$ cannot deviate from zero to more than $A(2k+3)L^{-\alpha/2}$ in this case, i.e. the direction of momentum is preserved with the accuracy $O(L^{-\alpha/2})$. Since the potential is of order $L^{-\alpha}$ in the region under consideration, the value of kinetic energy, and hence the absolute value of the momentum, is preserved with the accuracy $O(L^{-\alpha})$. Thus, both components of the momentum are preserved in this case with the accuracy $O(L^{-\alpha/2})$, as required. □

This lemma does not mean that the trajectories staying far from the boundary are uniformly close to straight lines (see figure 7). It rather says that the trajectory $(x(t), y(t))$ is confined within a narrow wedge around the ray $(x = x(0) + p_x(0)t, y = y(0) + p_y(0)t)_{t \geq 0}$.

It follows, in particular, that a trajectory which enters the region $kx-|y|\geq L$ with the values $(p_x,p_y)$ of momenta such that $kp_x>|p_y|+O(L^{-\alpha/2})$ will stay in this region forever and its distance to the boundary will grow without bound. The latter implies, by the above lemma, that the difference between the values of momenta at some time $t_0$ and at a larger time $t_1$ tends to zero as $t_0\to+\infty$, no matter how large $t_1$ is. Hence, $p(t)$ has a limit as $t\to+\infty$. Note that the convergence to the limit is locally uniform. The rate of convergence is determined by the speed with which the distance to the boundary grows, and the latter is proportional to the momentum, so for nearby trajectories the rate of convergence is approximately the same. Therefore, the limiting value $p(+\infty)$ depends indeed continuously on the initial conditions in this particular case.

The behavior on a *finite* distance from the boundary is easily understood at large $x$. Indeed, by the corner scaling assumption, at large $x$ and small $\varepsilon$ the system on a finite distance to the upper (or lower) boundary becomes close to the integrable one. Near the upper boundary the integrable limit is defined by the Hamiltonian

$$H=\frac{1}{2}(p_{\|}^2+p_{\perp}^2)+W_+(kx-y). \tag{9.2}$$

where $p_{\|}=\frac{1}{\sqrt{1+k^2}}(p_x+kp_y),p_{\perp}=\frac{1}{\sqrt{1+k^2}}(kp_x-p_y)$, and, near the lower boundary by the Hamiltonian:

$$H=\frac{1}{2}(p_{\|}^2+p_{\perp}^2)+W_-(kx+y). \tag{9.3}$$

where $p_{\|}=\frac{1}{\sqrt{1+k^2}}(p_x-kp_y),p_{\perp}=\frac{1}{\sqrt{1+k^2}}(kp_x+p_y)$. In both cases $p_{\|}$ is the constant of motion for the limit system. The behavior of $p_{\perp}$ is quite simple as well: it just grows monotonically, so the distance to the boundary (i.e. $(kx-y)$ in the case of upper boundary and $(kx+y)$ in the case of lower boundary) either grows all the time without bound or it decreases, first, to its minimal value where $p_{\perp}=0$ and then starts to increase. Note that for fixed values of $H$ and $p_{\|}$ the absolute value of $p_{\perp}$ is uniquely defined (via (9.2) or ( 9.3)) by the distance to the boundary.

Let $L$ be a fixed finite constant and let $M$ be sufficiently large. Define upper and lower boundary layers as $\{b_++kx-y\leq L,x\geq M\}$ and $\{-b_-+kx+y\leq L,x\geq M\}$. In the limit $M\to+\infty$, $\varepsilon\to0$ the system in the boundary layers limits to the integrable systems ( (9.2) or ( 9.3)), hence the following result holds

**Lemma 7.** *For any fixed $L$, sufficiently large $M$ and sufficiently small $\varepsilon$, any orbit of system (5.2) starting within one of the boundary layers with the energy $H=1$ must leave it in a finite time, bounded from above by some $\tau_{exit}(L)$ which is independent of the initial conditions. During the time spent within the boundary layer, the parallel momentum $p_{\|}$ is approximately preserved (i.e. it is preserved with the accuracy increased as $M\to+\infty$, $\varepsilon\to0$, uniformly with respect to the initial conditions inside the boundary layer), and the normal momentum grows monotonically. If the orbit does not enter or exit the boundary layer from the $x=M$ boundary, then $p_{\perp}(exit)\approx-p_{\perp}(entrance)$. Moreover, if such an orbit penetrates the boundary layer to the distance $r$, then $\frac{1}{2}p_{\perp}^2(entrance)\gtrsim W_{\pm}(r)-W_{\pm}(L)$.*

*Proof.* Just note that the same kind of behavior is shown by the limit integrable systems (9.2) or (9.3) (the approximate identities become exact, of course), and the orbits of the system (5.2) in the boundary layers are close to the orbits of (9.2) or (9.3) for any finite time, uniformly with respect to the initial conditions, provided $M$ is large and $\varepsilon$ is small. □

Notice that while the momenta obey, asymptotically, the billiard reflection laws, the actual Hamiltonian trajectory upon exiting the boundary layer may have a nonzero, yet finite (of order $\tau_{exit}(L)\cdot p_{||}$) shift in the coordinates $(x,y)$ in the direction parallel to the boundary.

Combining the results of the two lemmas above, we may now characterize the behavior of all trajectories of (5.2) at large $x$. As explained in section 3, for almost all initial conditions, a billiard trajectory starting in a corner domain with $p_x(0)\geq 0$ hits the boundaries finitely many times and then exits the corner region with some exit direction. We now show that the Hamiltonian trajectory at large $x$ has the same property:

**Lemma 8.** *Consider the rescaled system (5.2), satisfying the corner scaling assumption. Let $x(0), L$ be sufficiently large, $\varepsilon$ sufficiently small, and let $(x(0),y(0),p_x(0),p_y(0))\in\overline{C}_\varepsilon$, such that $\frac{p_y(0)}{p_x(0)}\neq\frac{y(0)}{x(0)}$ or $p_x(0)\geq 0$. Then, after a finite time the orbit does not visit the boundary layers, and the values of the momenta become $O(L^{-\alpha/2})$-close to the billiard exit direction.*

*Proof.* Consider a fan of billiard trajectories in the corner region $\overline{C}_\varepsilon$ (see 5.3) with the initial conditions $(x(0),y(0),p_x(0)+u,p_y(0)+v)$ where $(u,v)$ are small (of order $L^{-\alpha/2}$). The billiard trajectories all have a finite number of reflections, all of them occur at $x$ values larger than $Kx(0)$, for some constant $K$ depending on $\left|\frac{p_y(0)}{p_x(0)}-\frac{y(0)}{x(0)}\right|$. Furthermore, after a finite time $\tau$, independent on the value of $x(0)$, the momenta of the billiard trajectories will be close to the exit direction. From the two previous lemmas, it follows that the corresponding Hamiltonian trajectory stays (for any given finite interval of time, provided $x(0)$ was taken sufficiently large) on a finite distance in configuration space and a small distance in momenta space to this set of the billiard trajectories. Hence, if the exit direction is not parallel to either one of the two boundaries (see figure 8), it follows that for sufficiently large $L$ the momenta of the billiard trajectories at time $\tau$ are non-parallel to the boundaries as well (i.e. $kp_x(\tau)-|p_y(\tau)|\neq 0$), and the same is true for the momentum of the Hamiltonian trajectory. It follows then (by lemma 6) that the momentum of the Hamiltonian trajectory is approximately conserved for all $t\geq\tau$, i.e. the Hamiltonian trajectory remains close, in the above sense, to the fan of billiard trajectories for all time, proving the lemma for this case.

Now consider the case for which the exit direction is parallel to one of the boundaries, as in figure 9. Then, for $(u,v)=(0,0)$, the billiard trajectory satisfies for all $t>\tau$, $kp_x(t)-|p_y(t)|=0$. In this case, we have that the Hamiltonian trajectory is close to the fan of billiard trajectories for $t\leq\tau$, and $kp_x(\tau)-|p_y(\tau)|=O(L^{-\alpha/2})$. With no loss of generality, consider the case where $kp_x+p_y=O(L^{-\alpha/2})$, namely the Hamiltonian orbit is almost parallel to the lower boundary at $t=\tau$. By lemma 6, this estimate holds as long as the orbit stays outside of the boundary layers. So, its distance to the upper boundary will steadily grow, but the orbit may, in principle, enter the lower boundary layer. Let us prove that the estimate

$$kp_x+p_y=O(L^{-\alpha/2}) \tag{9.4}$$

will hold true for all times in this case as well.

Indeed, fix some $L'>L$ such that

$$W_-(L)\gg(L')^{-\alpha}, \tag{9.5}$$

and notice that by (5.5)

$$W_-(L)\gg W_-(L').$$

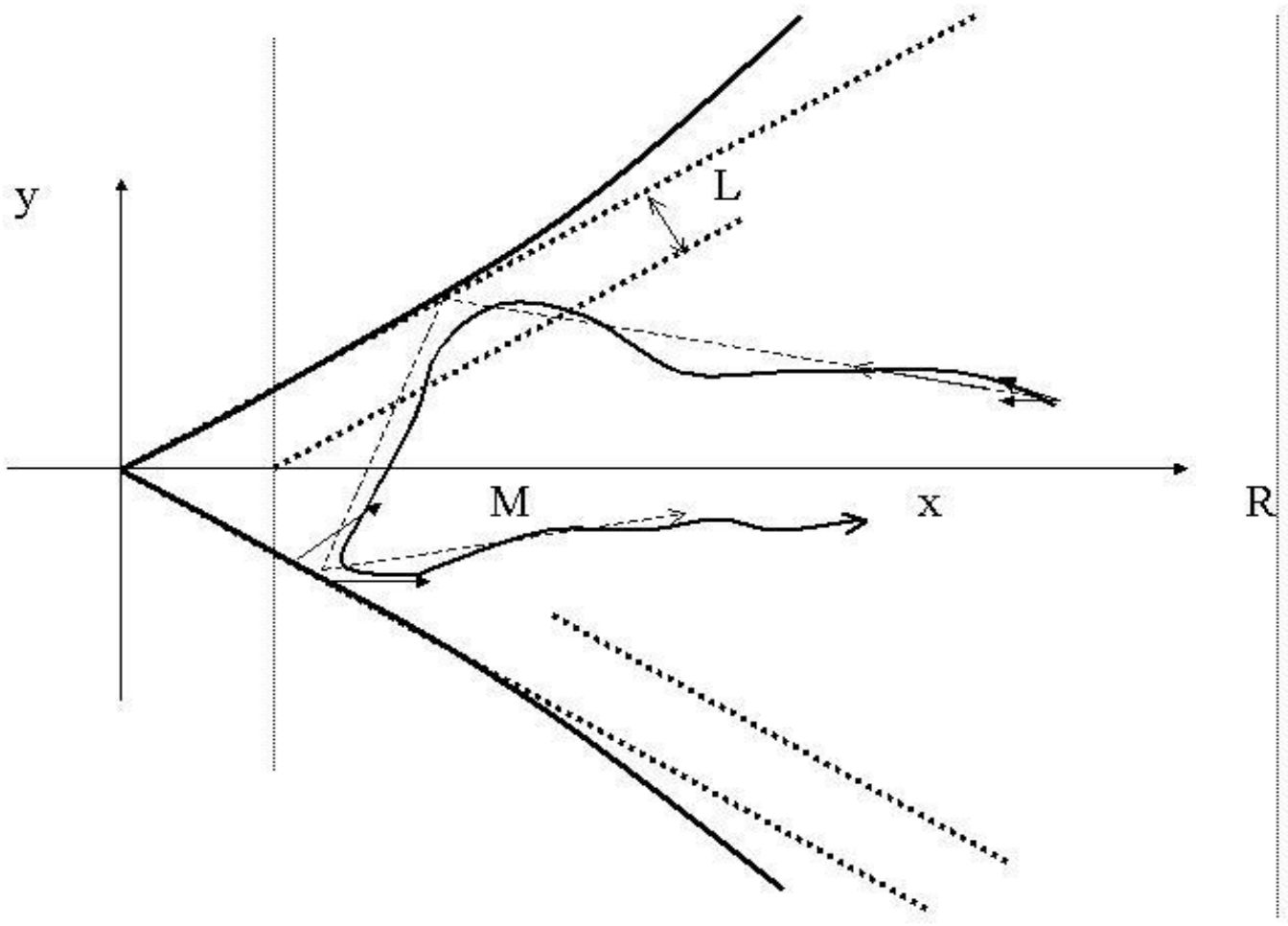

FIGURE 8. A trajectory which is not parallel to the boundary
and does not aim to the corner follows closely a ray of billiard trajectories.

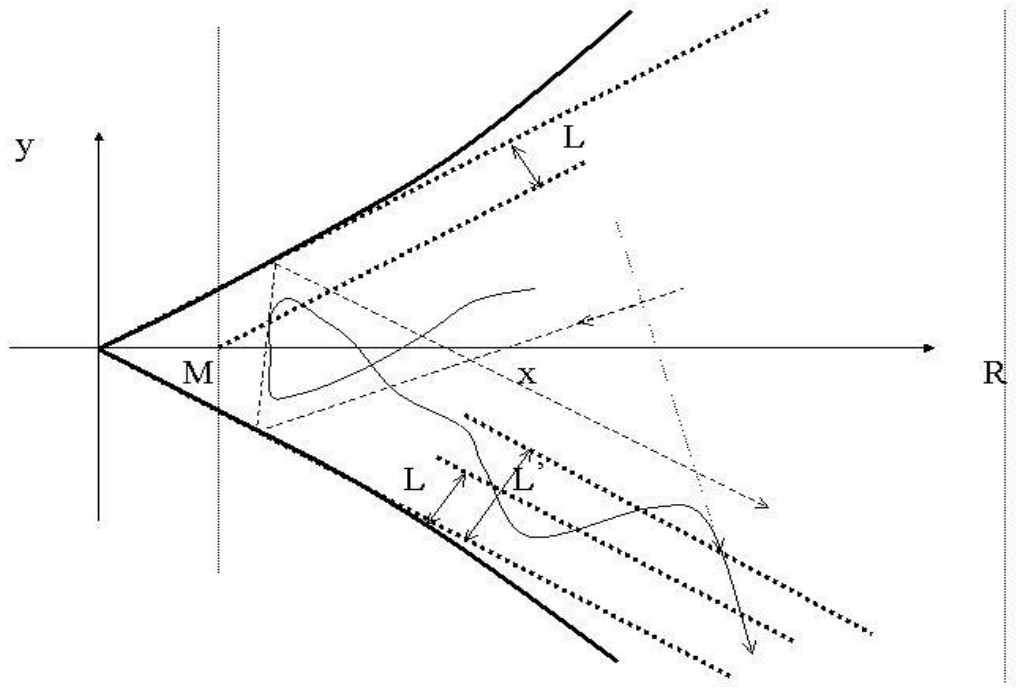

FIGURE 9. A trajectory which is parallel to the boundary
cannot re-enetr $L$, as shown, without hitting the upper boundary first (the dotted line).

If the orbit enters the lower boundary layer of size $L$, it must leave it, and then the larger boundary layer, of size $L'$, after the time $\tau_{exit}(L')$, due to lemma 7. During this time, the parallel component of the momentum is approximately conserved and the perpendicular component of the momentum is bounded by $O(L^{-\alpha/2})$ so we still have (9.4). To prove the lemma it remains to prove that after the orbit left the size $L'$ boundary layer, it can never enter the smaller, size $L$, boundary layer once again; we will have then (9.4) for all times, due to lemma 6.

First note that (9.4) holds, by lemma 6, as long as the orbit stays outside the size $L$ boundary layers. Therefore, the orbit cannot come close to the upper boundary until it visits the lower boundary layer of size $L$ at least one more time. Now, if upon exiting the size $L'$ lower boundary layer the orbit returns to it and then reaches the size $L$ lower boundary layer within, then (9.5) and lemma 7 imply that $\frac{1}{2}p_{\perp}^{2}(L'\ entrance) \geq W_{-}(L) - W_{-}(L') \gg (L')^{-\alpha}$. By lemma 6, this means that the same was true all the time the orbit stayed outside the size $L'$ boundary layers. Continuing the orbit in the backward time we see that it came from the upper boundary layer of size $L'$, i.e. it was there before entering the lower boundary layer of size $L$. The contradiction proves the claim. □

We see that for any outgoing orbit starting at sufficiently large $x$ the distance to the boundary must tend to infinity. By lemma 6, this implies that for every such orbit momenta must have a finite limit at $\varepsilon = 0$. Moreover, it follows from our proof that the distance to the boundary tends to infinity locally uniformly with respect to initial conditions and $\varepsilon$. Hence, the limit value, as $\varepsilon \to +0$ and $t \to +\infty$, depends on the initial conditions continuously. By reversibility, the same is valid as $t \to -\infty$. It remains to recall that by our scattering assumption all the trajectories must come to the region of sufficiently large $x$ both as $t \to +\infty$ and $t \to -\infty$. Now, applying the previous arguments, we have the proposition.

WEIERSTRASS INSTITUTE FOR APPLIED ANALYSIS AND STOCHASTICS,, MOHRENSTR 39, 10117 BERLIN, GERMANY., TURAEV@WIAS-BERLIN.DE

DEPARTMENT OF APPLIED MATHEMATICS AND COMPUTER SCIENCE, THE WEIZMANN INSTITUTE OF SCIENCE,, P.O.B. 26, REHOVOT 76100, ISRAEL., VERED@WISDOM.WEIZMANN.AC.IL